\documentclass[12pt]{iopart}
\pdfoutput=1

\usepackage{amsopn}   
\usepackage{amssymb}
\usepackage{epsfig}
\usepackage[T1]{fontenc}
\usepackage[utf8]{inputenc}
\usepackage{color}
\usepackage[normalem]{ulem}
\usepackage{tikz}
\usetikzlibrary{positioning}
\usepackage{tabularx}
\usepackage{enumitem}
\usepackage{graphicx}
\usepackage{caption}
\usepackage{subcaption}
\usepackage{ragged2e}
\DeclareCaptionJustification{justified}{\justifying}
\usepackage{dcolumn}
\usepackage{bm}
\usepackage{comment}
\usepackage{siunitx}
\usepackage[title]{appendix}
\usepackage{booktabs}
\usepackage{float}
\usepackage{multirow}
\usepackage{lipsum}
\usepackage{tabulary,xcolor}
\usepackage{amsfonts}
\usepackage{verbatim} 
\usepackage{url,morefloats,floatflt,cancel}
\usepackage{hyperref}
\usepackage{enumitem}
\usepackage{orcidlink}

\usepackage{iopams}  
    \usepackage[
    backend=bibtex,
   style=nature,
   doi=false, url=false,labeldate=year,date=year,
  ]{biblatex}

\long\def\textout#1{}

\begin{document}

\title[Phase transition properties via partition function zeros in the Blume-Capel model]{Phase transition properties via partition function zeros: The Blume-Capel ferromagnet revisited}

\author{Leïla Moueddene\textsuperscript{1,2,4$\star$}\orcidlink{0009-0001-7588-3835},
Nikolaos G Fytas\textsuperscript{3$\dagger$}\orcidlink{0000-0002-9428-1709},
Bertrand Berche\textsuperscript{1,4$\ddagger$}\orcidlink{0000-0002-4254-807X}}

\address{$^1$Laboratoire de Physique et Chimie Théoriques, CNRS - Université de Lorraine, UMR 7019,
Nancy, France}

\address{$^2$Centre for Fluid and Complex Systems, Coventry University, Coventry CV1 5FB, United Kingdom}

\address{$^3$School of Mathematics, Statistics and Actuarial Science, University of Essex, Colchester CO4 3SQ, United Kingdom}

\address{$^4$L4 collaboration, Leipzig-Lorraine-Lviv-Coventry, Europe}

\begin{center}
    $\star$ \href{mailto:leila.moueddene@univ-lorraine.fr}{\small leila.moueddene@univ-lorraine.fr}\,\\
    $\dagger$ \href{mailto:nikolaos.fytas@essex.ac.uk}{\small nikolaos.fytas@essex.ac.uk}
    \\ $\ddagger$ \href{mailto:bertrand.berche@univ-lorraine.fr}{\small bertrand.berche@univ-lorraine.fr}
\end{center}

\date{\today}

\begin{abstract}
Since the landmark work of Lee and Yang, locating the zeros of the partition function in the complex magnetic-field plane has become a powerful method for studying phase transitions. Fisher later extended this approach to complex temperatures, and subsequent generalizations introduced other control parameters, such as the crystal field. In previous works [Moueddene \emph{et al}, J. Stat. Mech. (2024) 023206; Phys. Rev. E {\bf 110} (2024) 064144] we applied this framework to the two- and three-dimensional Blume–Capel model--a system with a rich phase structure where a second-order critical line meets a first-order line at a tricritical point. We showed that the scaling of Lee–Yang, Fisher, and crystal-field zeros yields accurate critical exponents even for modest lattice sizes. In the present study, we extend this analysis and demonstrate that simulations need not be performed exactly at the nominal transition point to obtain reliable exponent estimates. Strikingly, small system sizes are sufficient, which not only improves methodological efficiency but also advances the broader goal of reducing the carbon footprint of large-scale computational studies.
\end{abstract}

\submitto{Journal of Statistical Mechanics}

\maketitle

\section{Introduction}
\label{sec:Intro}

\begin{figure}[ht]
    \centering
\includegraphics[width=0.70\textwidth]{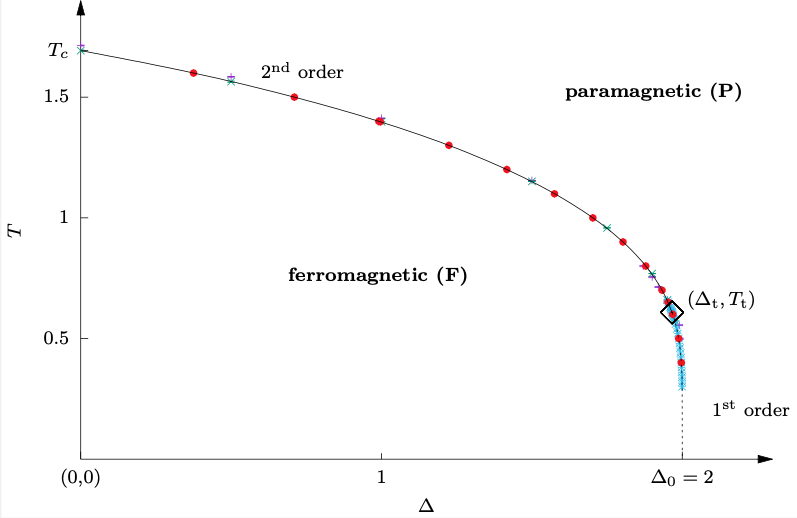} \caption{Phase diagram of the 2D Blume-Capel model as reported in the literature. Several sets of results are shown from~\cite{silva,malakis,kwak,Zierenberg}, as adapted from~\cite{Zierenberg}.}
\label{fig:phasediaggeneral} 
\end{figure}

In this work, we build upon the extensive studies of the critical and tricritical properties of the two-dimensional (2D) Blume-Capel model~\cite{blume_theory_1966,capel_possibility_1966}. This model generalizes the Ising model by allowing spins to take values $\sigma_i = \pm 1, 0$ instead of just $\pm 1$. Along with the usual nearest-neighbor ferromagnetic exchange interaction $J$ and coupling to an external magnetic field $H$, the Blume-Capel model introduces a crystal-field interaction $\Delta$. The latter acts as a chemical potential that controls the density of zero-spin states: larger values of $\Delta$ favour configurations where $\sigma_i=0$ is more prevalent. The introduction of the uncoupled zero-spin state, like invisible states in the Potts model \cite{invisible} that only contribute to the entropy, enriches the phase diagram, giving rise to both second-order and first-order phase transitions that meet at a tricritical point, see figure~\ref{fig:phasediaggeneral}. This added complexity makes the Blume-Capel model a particularly valuable framework for investigating multicritical phenomena in statistical physics~\cite{Lawrie, bernasconi_critical_1971,fytas11,malakis12,selke-10,fytas-selke13,fytas18,vatansever20,macedo24,mataragkas25,mataragkas25b,bisson25}.

Since it was first proposed, the Blume-Capel model has been investigated through mean-field theory, perturbative expansions, and numerical simulations on a variety of lattices, primarily in two and three dimensions; see, e.g.,~\cite{fytas11,fytas2012,fytas2013,zierenberg2015}. The vast majority of studies have focused on the square lattice, employing a broad range of methods. These include real-space renormalisation~\cite{berker1976rg}, Monte Carlo simulations and Monte Carlo renormalisation-group approaches~\cite{landau1972,kaufman1981,selke1983,selke1984,landau1986,xavier1998,deng2005,silva,hurt2007,malakis,kwak,Zierenberg}, $\epsilon$-expansion techniques~\cite{stephen1973,chang1974,tuthill1975,wegner1975}, high- and low-temperature series expansions~\cite{fox1973,camp1975,burkhardt1976}, and transfer-matrix calculations~\cite{beale1986,kim17,xavier1998,qian05,blote19}.

Here, we extend our analysis of partition function zeros--a technique we previously showed to be highly effective for probing critical properties of the 2D Blume-Capel model, even with very modest system sizes~\cite{moueddene_critical_2024} (this constraint was also applied in three dimensions in~\cite{e26030221,moueddene_critical_2024-1} and on complete graphs in~\cite{LTP}). Our strategy, detailed in~\cite{moueddenethesis}, remains unchanged: restrict simulations to extremely small lattices and extract maximal information from the zeros of the partition function after analytic continuation into the complex plane of the magnetic field (Lee–Yang zeros), temperature (Fisher zeros), or the crystal-field parameter.

Throughout this study, we adopt the same notation as in the referenced article~\cite{moueddene_critical_2024}, to which we refer readers for detailed definitions.
The Hamiltonian of the Blume-Capel model is given by~\cite{blume_theory_1966,capel_possibility_1966}
\begin{equation}
{\cal H}
= -J \sum_{(i,j)} \sigma_i \sigma_j
- H \sum_i \sigma_i
+ \Delta \sum_i \sigma_i^2
= E_J - H M + \Delta E_\Delta ,
\label{eq_HBCModel}
\end{equation}
and the corresponding partition function for a system of $N$ spins in $d$ dimensions on a simple hypercubic lattice can be expressed in terms of the density of states $\rho(E_J,M,E_{\Delta})$ as
\begin{equation}
Z_N(\beta J, \beta H, \beta \Delta)
= \sum_{E_J=-dN}^{dN}
\sum_{M=-N}^{N}
\sum_{E_{\Delta}=0}^{N}
\rho(E_J,M,E_\Delta), x^{E_J} y^M z^{E_{\Delta}} ,
\end{equation}
where $x = e^{-\beta J}$, $y = e^{\beta H}$, and $z = e^{-\beta \Delta}$.

To illustrate the method, let us consider the Lee-Yang approach. We fix the temperature and crystal field at their critical values, $\beta_{\rm c}$ and $\Delta_{\rm c}$---for instance, along the second-order transition line or at the tricritical point—and examine the dependence of the partition function on the magnetic field. For a finite system, factoring out the smallest power of $y = e^{-\beta_{\rm c} H}$ transforms $Z_N(\beta_{\rm c} J,\, \beta_{\rm c} H,\, \beta_{\rm c} \Delta_{\rm c})$
into a polynomial in $y$. Extending $H$ to complex values, this polynomial has roots $H^{(j)}$, ordered by increasing imaginary part. At each zero $H^{(j)}$, the free-energy density,  
$-\beta_{\rm c}^{-1}\ln Z_N(\beta_{\rm c} J,\, H^{(j)},\, \beta_{\rm c} \Delta_{\rm c})$,
is singular. In the thermodynamic limit, these singularities manifest at real parameter values: the lowest zeros condense onto a continuous curve that pinches the real axis precisely at the critical point, $H_{\rm c} = 0$. In the Lee-Yang case, symmetry under $M \to -M$ further constrains all zeros to lie on the imaginary axis (in the Lee–Yang theorem, the $\mathbb{Z}_2$ symmetry ($M\to -M$) ensures that zeros come in reciprocal pairs, while ferromagnetic interactions further constrain all zeros of the partition function--in the fugacity $y=e^{-\beta H}$--to lie exactly on the unit circle, i.e., on the imaginary axis in the complex magnetic field plane. The validity of the theorem can nevertheless be compromised in some cases~\cite{Krasnytska_2015}).

At a critical (or tricritical) point, the finite-size scaling of the lowest zeros follows, to leading order, a power law:
\begin{equation}
    \operatorname{Im}\bigl(H^{(j)}\bigr) \sim L^{-y_h},
\end{equation}
where $y_h$ is the magnetic renormalisation-group exponent for the universality class considered. 

The extension to Fisher zeros and crystal-field zeros requires accounting for the fact that the real parts converge to non-zero values of the corresponding parameters:
\begin{eqnarray}
    &\operatorname{Im}(\beta^{(j)})\sim L^{-y_t},\qquad &|\operatorname{Re}(\beta^{(j)})-\beta_{\rm c}|\sim L^{-y_t},\label{EqFitBetaCritical}\\
    &\operatorname{Im}(\Delta^{(j)})\sim L^{-y_g},\qquad &|\operatorname{Re}(\Delta^{(j)})-\Delta_{\rm c}|\sim L^{-y_g}.
\end{eqnarray} Here, $y_t$ and $y_g$ are associated with even scaling fields (the smaller of which corresponds to a correction to scaling), while $y_h$ corresponds to the magnetic sector, i.e., an odd scaling field under the $\mathbb{Z}_2$ transformation $\forall i, \sigma_i \to -\sigma_i$. 
The renormalisation-group eigenvalues along the second-order line of the Ising model (IM)'s universality class are denoted $y_i^{\rm IM}$, whereas those at the tricritical point (TP) are denoted $y_i^{\rm TP}$. The known values in two dimensions are:
\begin{eqnarray}
    &&y_t^{\rm IM}=1,\quad y_h^{\rm IM}=\frac{15}{8},\\
    &&y_t^{\rm TP}=\frac{9}{5},\quad y_h^{\rm TP}=\frac{77}{40},\quad y_g^{\rm TP}=\frac{4}{5}.
\end{eqnarray}
These values serve as references, and we will demonstrate that they can be accurately recovered even using simulations of very small system sizes.

In the following, we explore various regions of the 2D Blume-Capel phase diagram, shown in figure~\ref{fig:phasediaggeneral}. 
For clarity, we will return to this phase diagram at several points throughout the article, each time explicitly highlighting the region under investigation.

\section{Numerical  observations along the transition lines of the Blume-Capel model}
\label{sec:num_obs}

In this section, we visualize representative spin configurations at various points in the phase diagram.

\begin{figure}[H]
    \centering
     \includegraphics[width=0.5\textwidth]{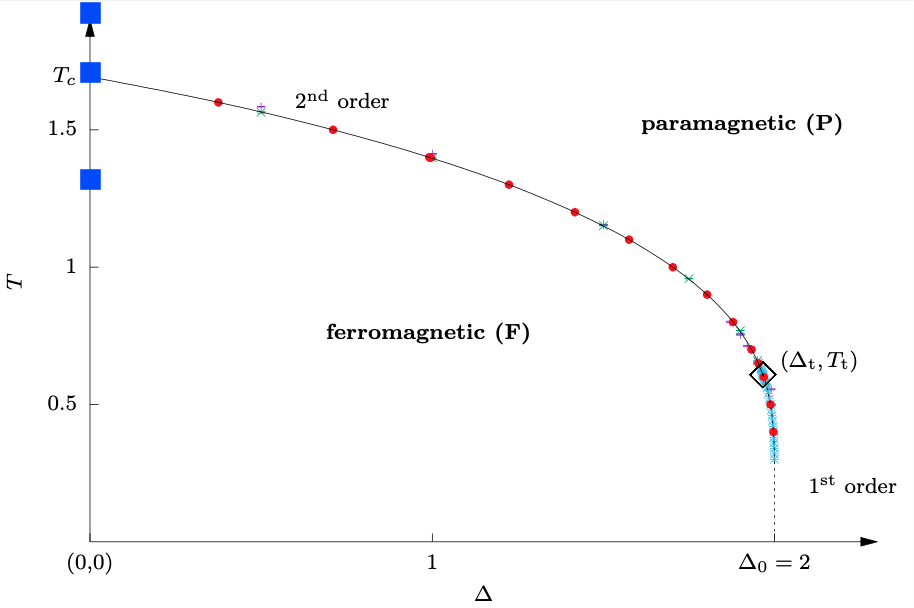}
     \includegraphics[width=0.40\textwidth]{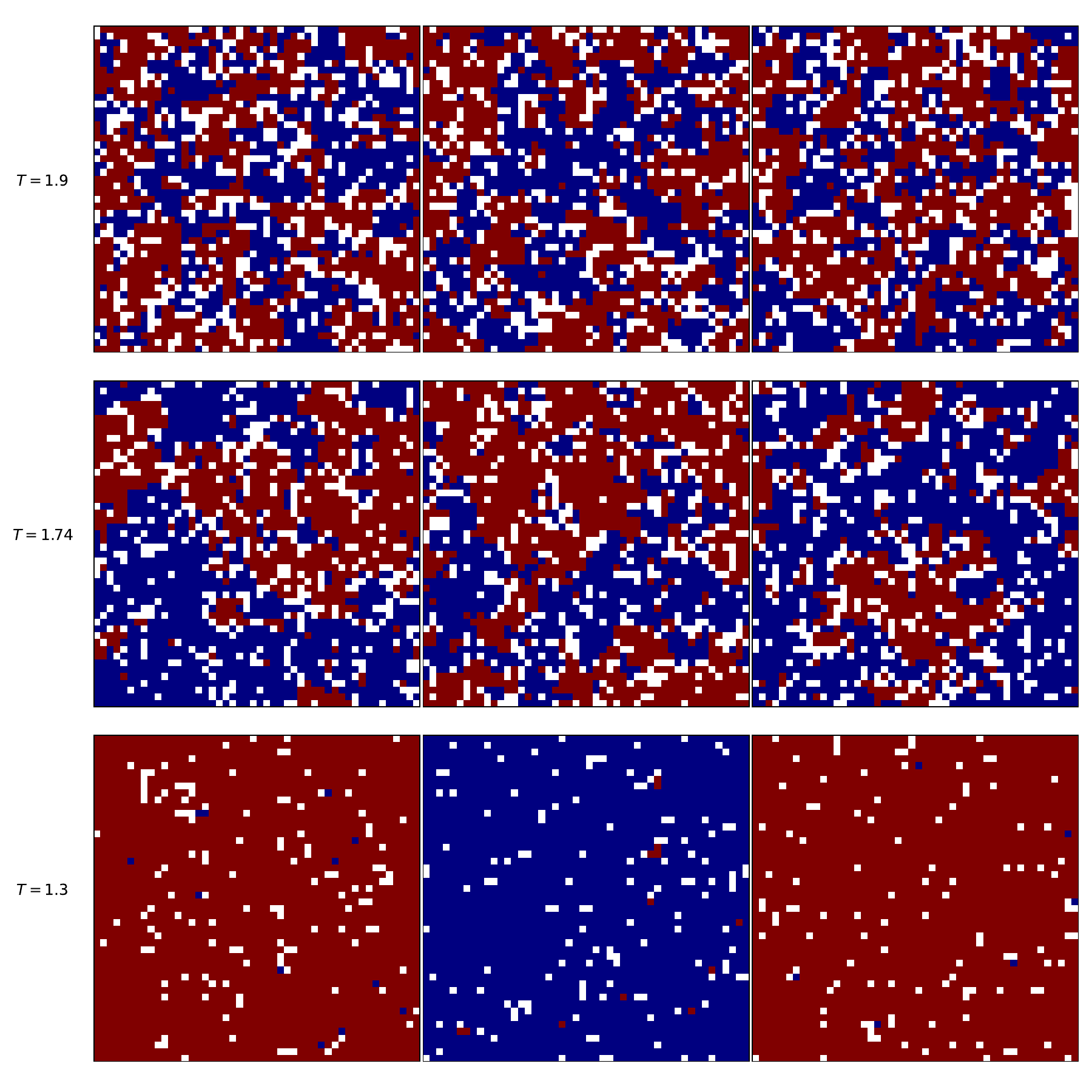}
     \caption{\textbf{Left panel:} Simulation points in the phase diagram. \textbf{Right panel:} Typical spin configurations at $\Delta=0$ (spins $+1$ in red, spins $-1$ in blue, and spins $0$ in white). Top: above $T_{\rm c}$; middle: at the Ising-model transition $T_{\rm c}$; bottom: below $T_{\rm c}$.}
\label{fig:2Dconfigs-L48} 
\end{figure}

Along the line $\Delta = 0$ (figure~\ref{fig:2Dconfigs-L48}), the system exhibits a clear evolution with temperature. At low temperatures, it is in a ferromagnetic, ordered phase dominated by either $+1$ spins (red online) or $-1$ spins (blue). As the temperature approaches the Ising transition point, clusters of red and blue spins become increasingly fragmented and fractal-like. Neither spin orientation dominates, and the overall magnetization approaches zero. Above the transition, in the paramagnetic regime, small red and blue clusters appear and average out, while zero spins (white) remain rare throughout this region. 

\begin{figure}[H]
    \centering
    \includegraphics[width=0.5\textwidth]{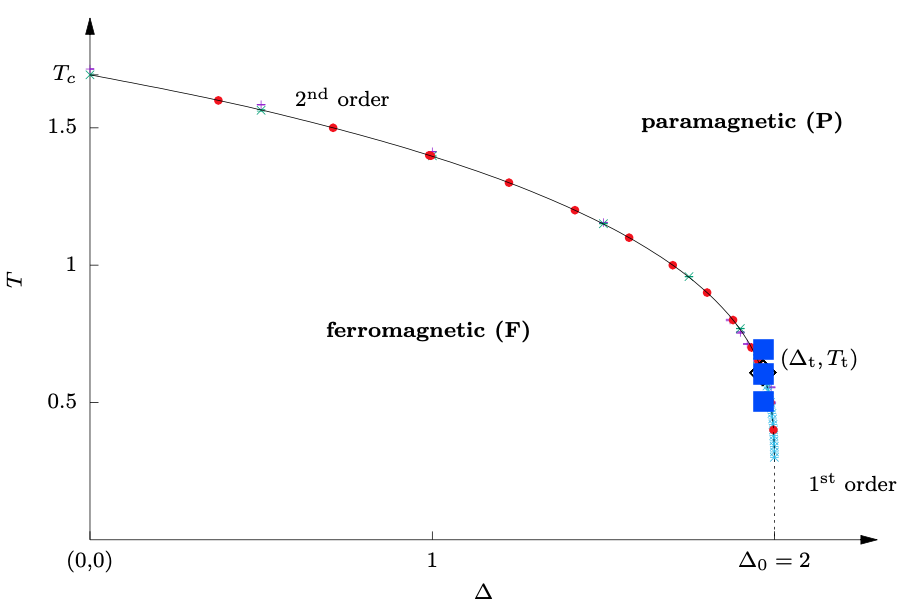}
    \includegraphics[width=0.40\textwidth]{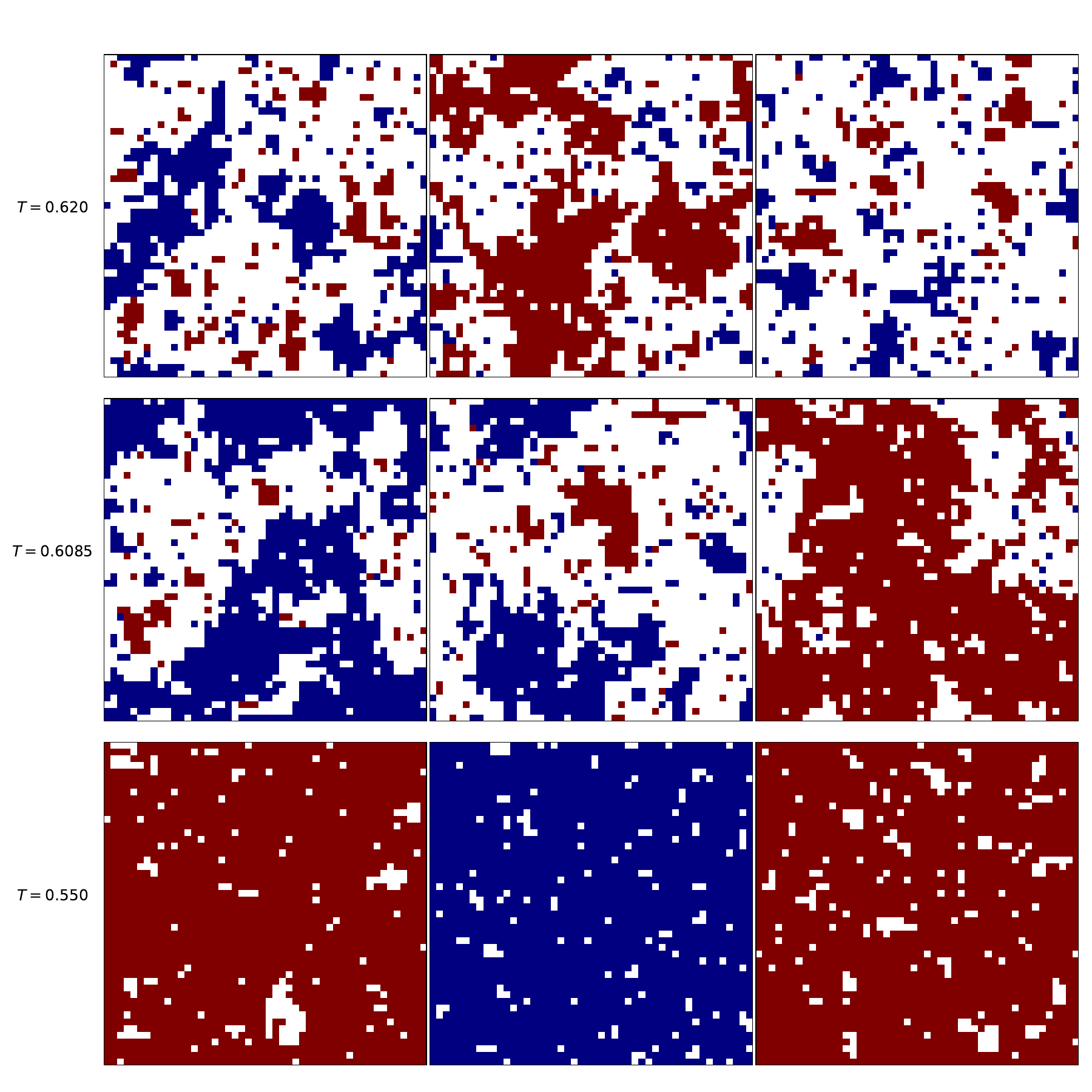}
    \caption{Typical spin configurations near the tricritical point (spins $+1$ in red, spins $-1$ in blue, and spins $0$ in white). Top: above $T_{\rm t}$; middle: at tricriticality; bottom: below $T_{\rm t}$. At tricriticality ($T = T_{\rm t}$), spin clusters appear fragmented and interpenetrating.}
\label{fig:2Dconfigs-L48tri} 
\end{figure}

As the crystal-field parameter $\Delta$ increases, white (zero-spin) sites become progressively more common, and white clusters start to appear in the configurations. Near the tricritical point (figure~\ref{fig:2Dconfigs-L48tri}), the ordered phase can be predominantly red, blue, or white, with white clusters often dominating, particularly close to the tricritical temperature. Even in the surrounding paramagnetic region, white spins remain abundant, highlighting the growing influence of the crystal field.

\begin{figure}[ht]
    \centering
      \includegraphics[width=0.5\textwidth]{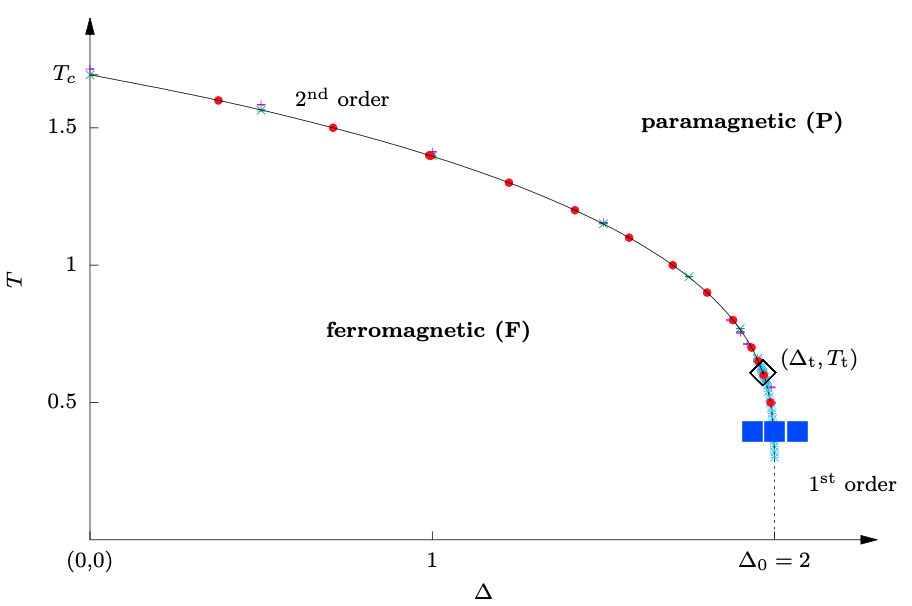}
\includegraphics[width=0.40\textwidth]{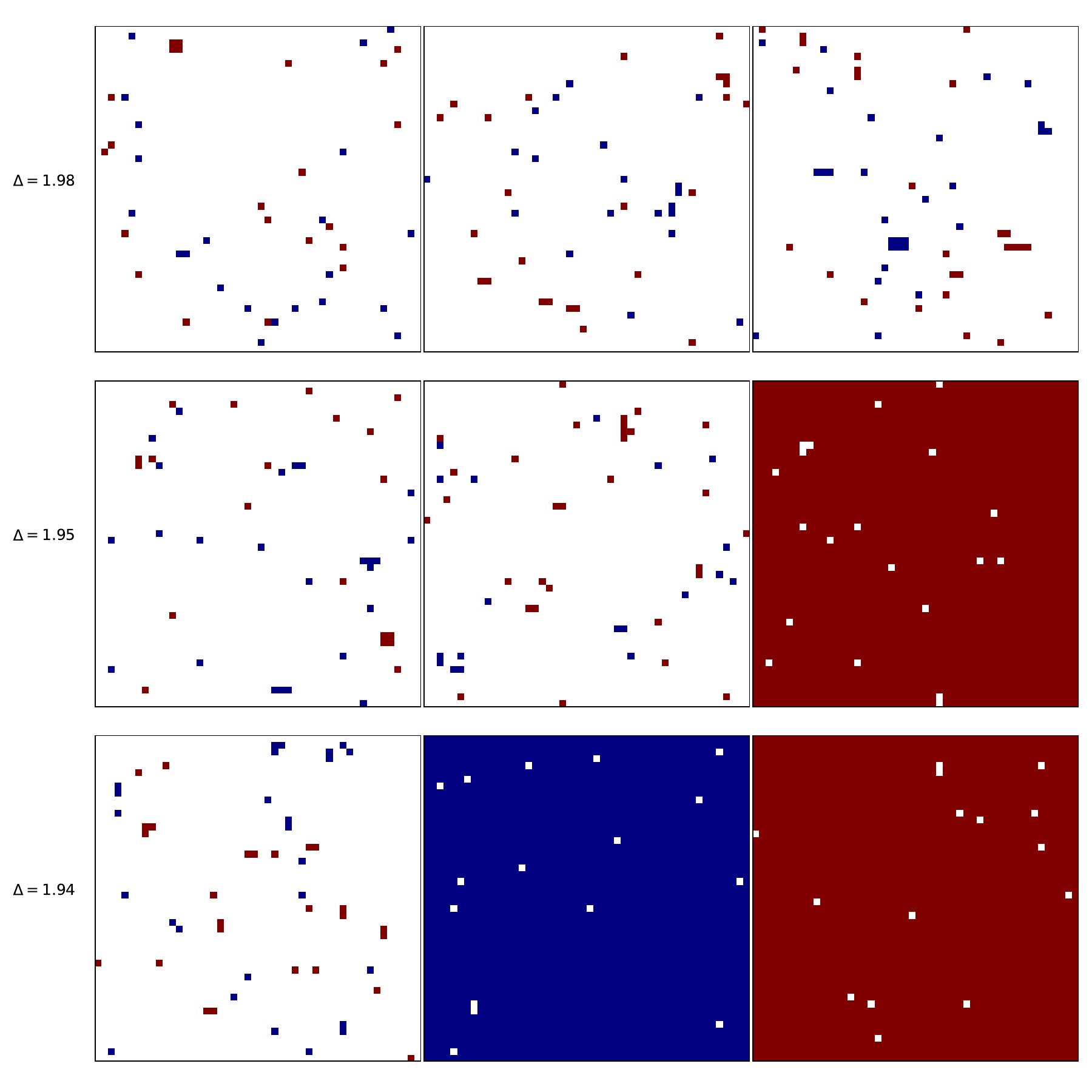}   
     \caption{Typical spin configurations near the first-order transition (spins $+1$ in red, spins $-1$ in blue, and spins $0$ in white). At a temperature below $T_{\rm t}$: top, $\Delta$ above the transition; center, at the transition; bottom, in the ordered phase. Note that for a finite system the transition is rounded, and one of the ordered configurations may appear dominated by spins that do not contribute to the magnetization, resembling a disordered phase.}
\label{fig:2Dconfigs-L481st} 
\end{figure}

In the first-order transition regime (figure~\ref{fig:2Dconfigs-L481st}), the change is much sharper. In the paramagnetic phase, zero spins overwhelmingly occupy the lattice, whereas in the ordered phase, red or blue spins form large, well-defined clusters that dominate the system. This dramatic change in spin composition and structure illustrates the distinct character of first-order transitions. However, since the simulations are performed on finite lattices, the transitions appear slightly rounded and are not as abrupt as in the thermodynamic limit.

\section{Ising universality class}
\label{sec:Ising}

\subsection{Renormalisation-group scaling dimensions along the critical line}
\label{sec:RG_scaling}

Although the universality class of the 2D Blume-Capel model has been extensively studied--as illustrated by the phase diagram presented in the introduction--we begin here by analyzing the scaling behaviour of three types of zeros: Fisher zeros in the complex-temperature plane, zeros in the crystal-field plane, and Lee-Yang zeros in the complex magnetic-field plane. In this section, we focus solely on the finite-size scaling of the imaginary part of the first zero at four distinct points along the critical line.

The simulations cover system sizes $L = 16$ to $64$. This allows us to (1) confirm that all four points belong to the same universality class--namely, the 2D Ising class, with $y_t^{\rm IM} = 1$ and $y_h^{\rm IM} = 15/8$--and (2) validate our zero-based analysis approach, which is developed further throughout this work.

\begin{figure}[t]
    \centering
     \includegraphics[width=0.5\textwidth]{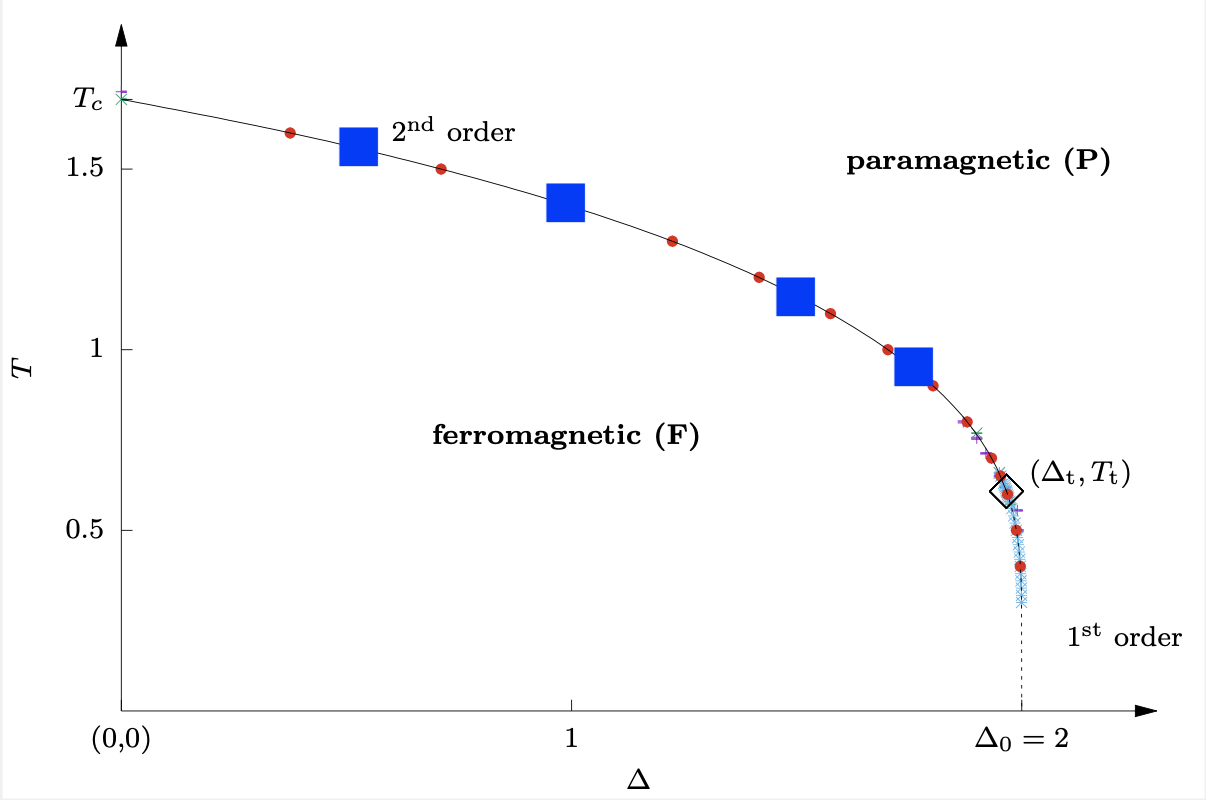} \includegraphics[width=0.45\textwidth]{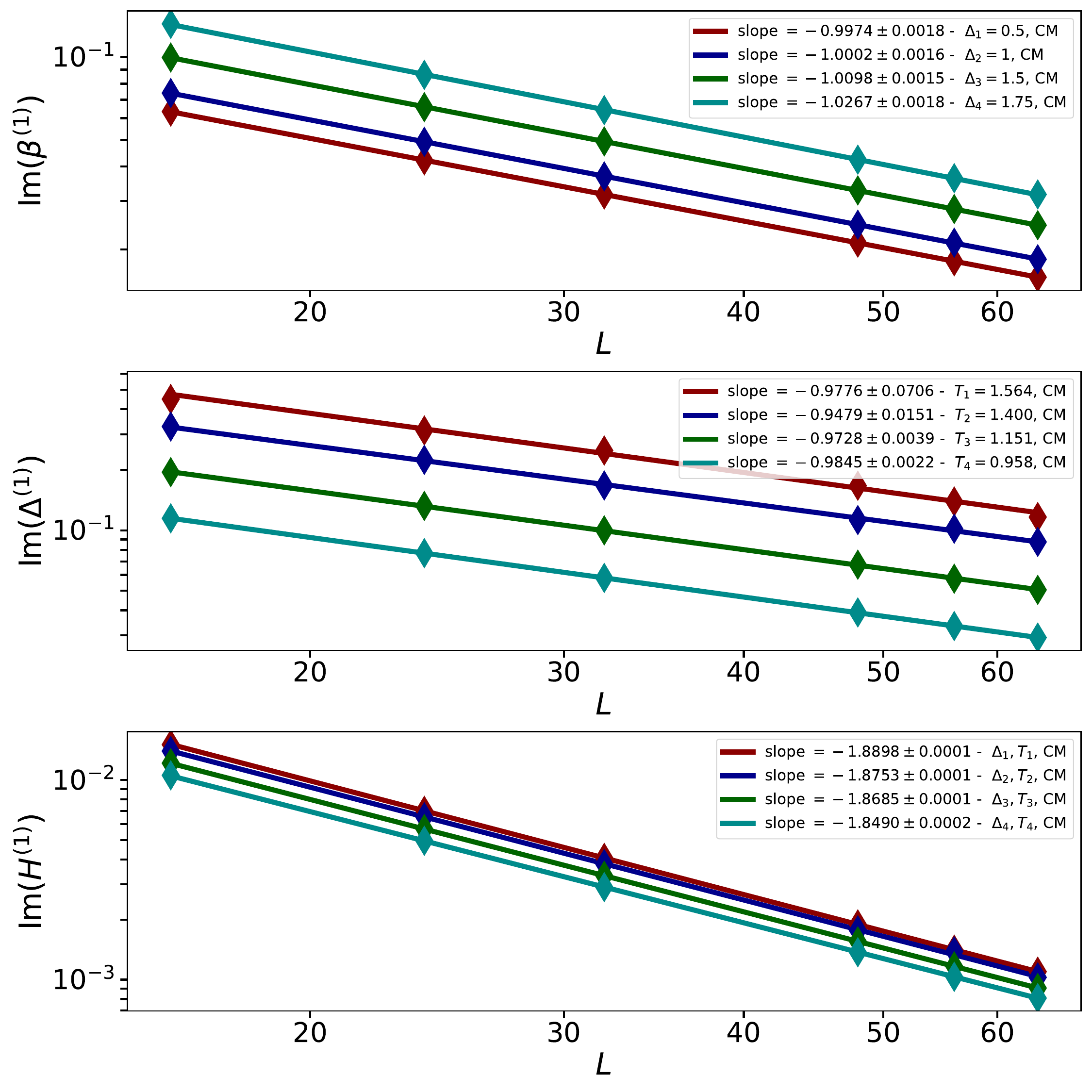}
     \caption{Finite-size scaling behaviour of the first zeros along the second-order line of the 2D Blume-Capel model. In the first two panels, the expected Ising-model renormalisation-group exponent is $y_t^{\rm IM}=1$, while in the third panel, $y_h^{\rm IM}=1.875$. CM denotes the cumulant method used to extract the zeros, which is discussed in detail below.}
\label{fig:phasediagzeros} 
\end{figure}

In figure~\ref{fig:phasediagzeros} (left panel), we show the locations in the phase diagram where the simulations were performed, with coordinates taken from the literature. The three panels below illustrate the finite-size scaling of the imaginary parts of the first Fisher zero, the first crystal-field zero, and the first Lee–Yang zero. The first two confirm $y_t^{\rm IM} \approx 1.000$, while the third confirms $y_h^{\rm IM} \approx 1.875$, in excellent agreement with the 2D Ising universality class. Detailed numerical data are reported in table~\ref{tab:widgets}. We note that the simulation at $\Delta = 1$ yields slightly lower-quality fits, which we attribute to the less precise knowledge of the critical line at this point; accordingly, we tuned $\Delta$ to $0.991$ for this simulation.

\begin{table}[H]
\centering
\begingroup\small
\begin{tabular}{|l|r|r|r|}
\hline
$\Delta$ & $T$ & $y_h^{\rm IM}$ & $\chi_s^2$ \\\hline
0.5  & 1.5640 & $1.8898(1)$ &   207.05 \\
&1.5657 & $1.8764(1)$ & $0.18$ \\ 
  & 1.5658 &1.8756(1) & 0.25 \\
 & \bf 1.5659 & \bf  1.8748(1) &\bf  1.77 \\ \hline
 
 0.991 & 1.400  & 1.8822(1) & 55.57 \\
  & 1.4005& 1.8779(1) & 8.33 \\ 
   & 1.4006 &  1.8770(1)& 3.89  \\ 
  & \bf   1.4007 & \bf  1.8761(1) & \bf  1.16 \\   \hline

1.5 & 1.1510&   1.8685(1)     & 19.33 \\
& 1.1506&  1.8725(1)    & 3.58 \\
& \bf 1.1504 & \bf  1.8745(1)      &\bf  1.91\\
& 1.1503 & 1.8755(1) &2.57\\ \hline
1.75  & 0.9580 &1.8490(2) & 132.62  \\ 
& 0.9558 & 1.8758(2) & 2.19 \\
& \bf 0.9559 &\bf 1.8746(2) & \bf 1.00 \\
& 0.9560 & 1.8735(2) & 0.42\\
& 0.9561 & 1.8723(2) & 0.45\\
\hline
\end{tabular}
\endgroup
\caption{\label{tab:widgets} Expected value $y_h^{\rm IM} = 1.875$. Best fits (in terms of $\chi^2_s$) are highlighted in bold. For $\Delta=0.991$, the $L=56$ data were excluded due to poor quality.}
\end{table}

\subsection{Role of off-critical observations}
\label{sec:off-crit_obs}

In this section, we revisit a key insight by Deger \emph{et al}~\cite{deger_determination_2019, deger_lee-yang_2020-1}, namely that critical properties can be accurately extracted without restricting simulations to the immediate vicinity of the transition. This is noteworthy, since finite-size studies typically focus on points of maximal divergence--such as susceptibility peaks--close to the critical point.

\begin{figure}[H]
    \centering
\includegraphics[width=0.45\textwidth]{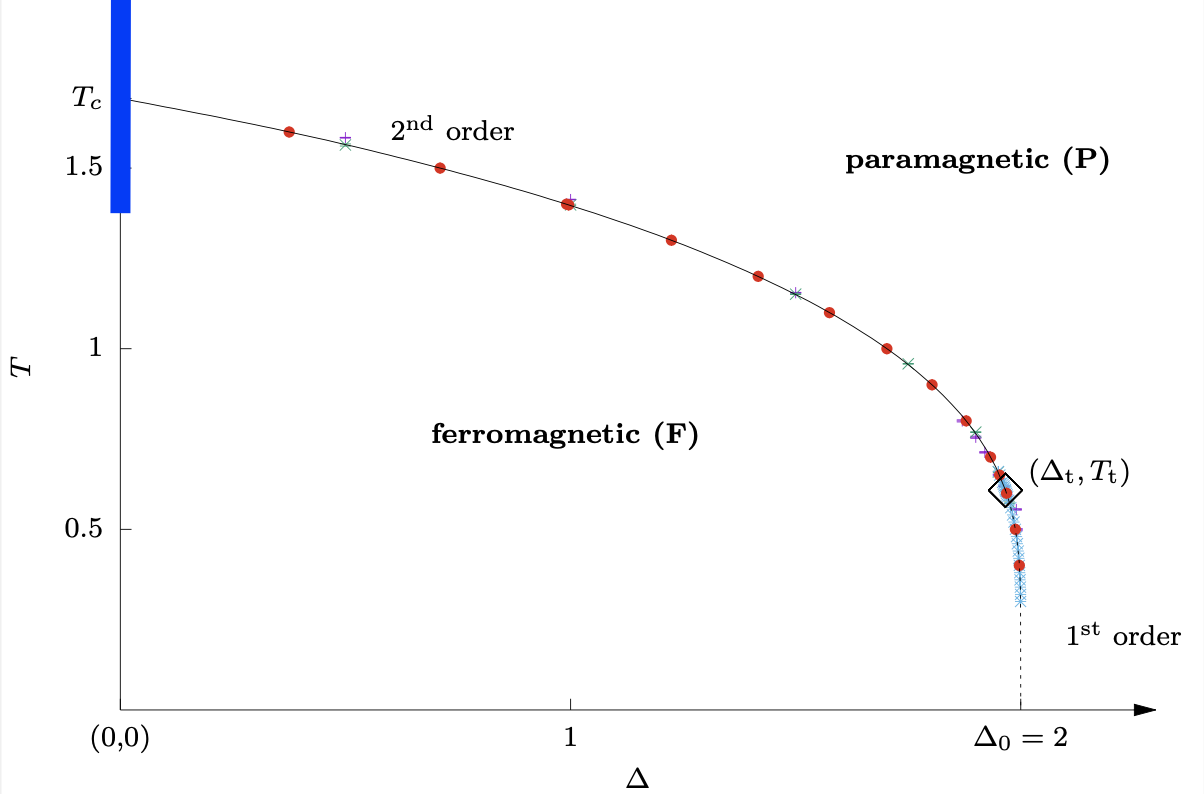}
\includegraphics[width=0.45\textwidth]{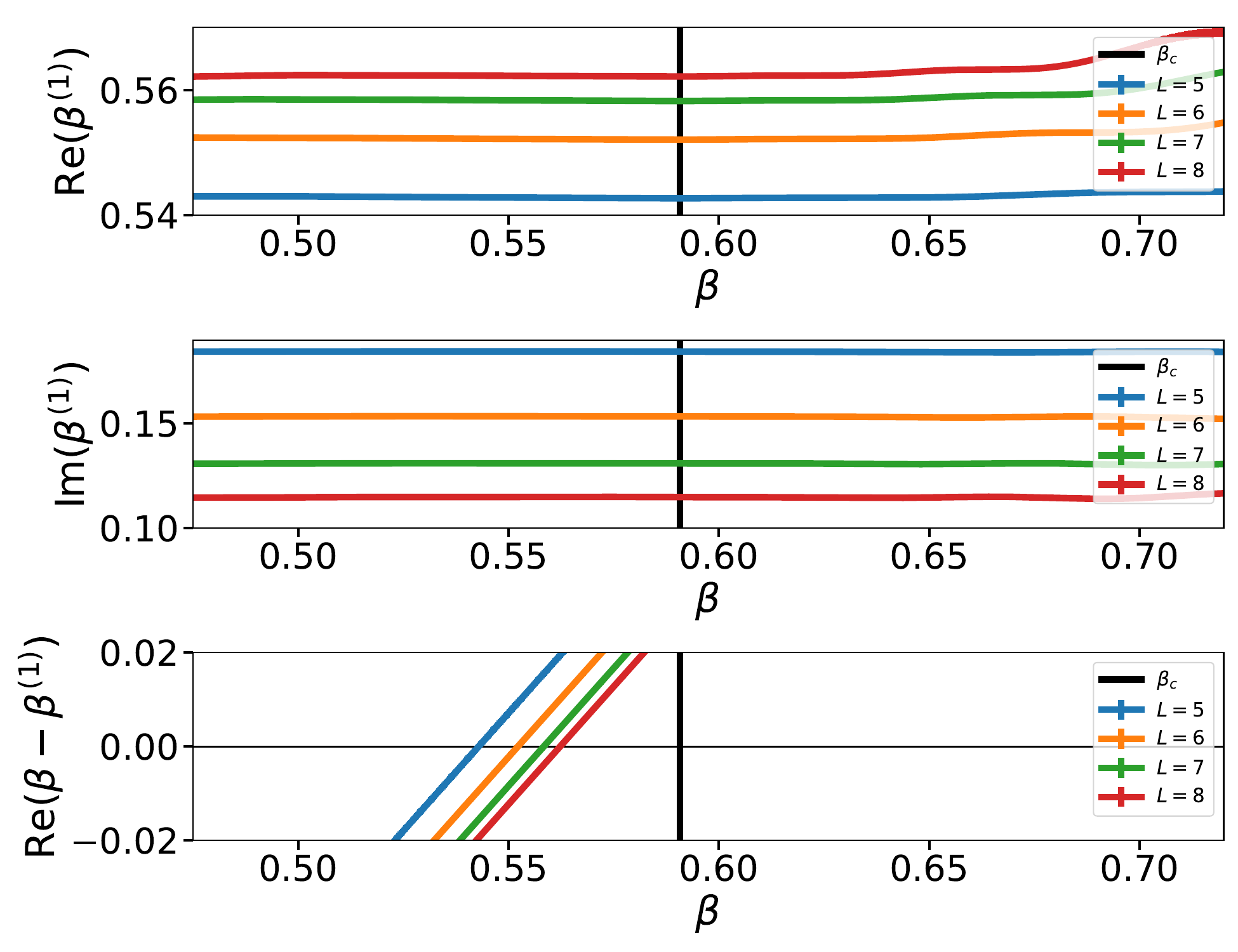}
\caption{Variation of the real and imaginary parts of $\beta^{(1)}$, obtained from the cumulant method, as a function of $\beta$. System sizes: $L = 5 - 8$.}
\label{fig:allureFisher}
\end{figure}

To illustrate this, we return to simulations at $\Delta = 0$ and examine the first Fisher zero across a broad temperature range, employing histogram reweighting to scan values of $\beta$. 
For each real $\beta$, we plot $\operatorname{Re} (\beta - \beta^{(1)})$ and identify the temperature where this quantity crosses zero. Around this crossing, both the real and imaginary parts of the first Fisher zero, $\operatorname{Re} (\beta^{(1)})$ and $\operatorname{Im} (\beta^{(1)})$, exhibit plateaus, remaining essentially constant. This behaviour is illustrated in figure~\ref{fig:allureFisher}, where the left panel shows the wide temperature interval explored through histogram reweighting in the phase diagram.  

\begin{figure}[H]
    \centering
     \includegraphics[width=0.6\textwidth]{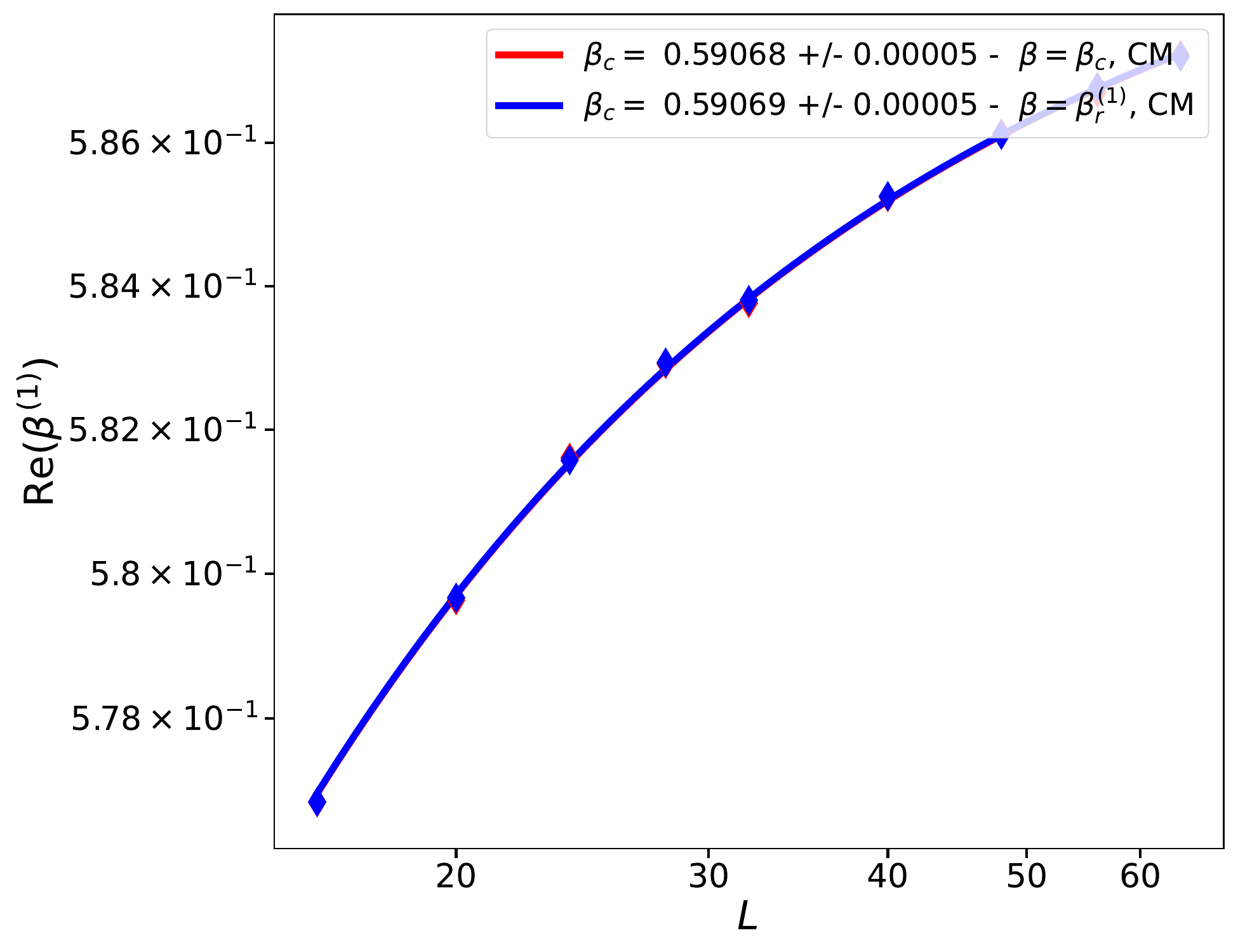}
     \caption{Finite-size scaling of the real part of the first Fisher zero, $\beta^{(1)}$, obtained via the cumulant method (CM) at $\beta = \beta_{\rm c}$ and $\beta = \beta_r^{(1)}$ to extract $\beta_{\rm c}$. System sizes: $L = 16$–$64$.}
\label{fig:FSSREb}
\end{figure}

The emergence of such plateaus explains why finite-size scaling of the Fisher zero can be reliably performed over a wide window, $0.80\,\beta_{\rm c} < \beta < 1.20\,\beta_{\rm c}$, using the cumulant method, even for very small systems ($L = 5 - 8$). This robustness allows critical exponents to be extracted from data collected far from the nominal transition, thereby confirming the conclusion of Deger \emph{et al}~\cite{deger_determination_2019, deger_lee-yang_2020-1}: critical exponents can be accurately determined even when the system is away from the phase transition.  
An application of this result is presented in figure~\ref{fig:FSSREb}, where we fit $\operatorname{Re}(\beta^{(1)})$ versus the system size $L$ to equation~(\ref{EqFitBetaCritical}) at two different temperatures, $\beta_{\rm c}$ and $\beta_r^{(1)}$. Both fits extrapolate to the same critical value $\beta_{\rm c}$, with a relative difference smaller than $0.009\%$.

\begin{figure}[t]
    \centering
     \includegraphics[width=0.6\textwidth]{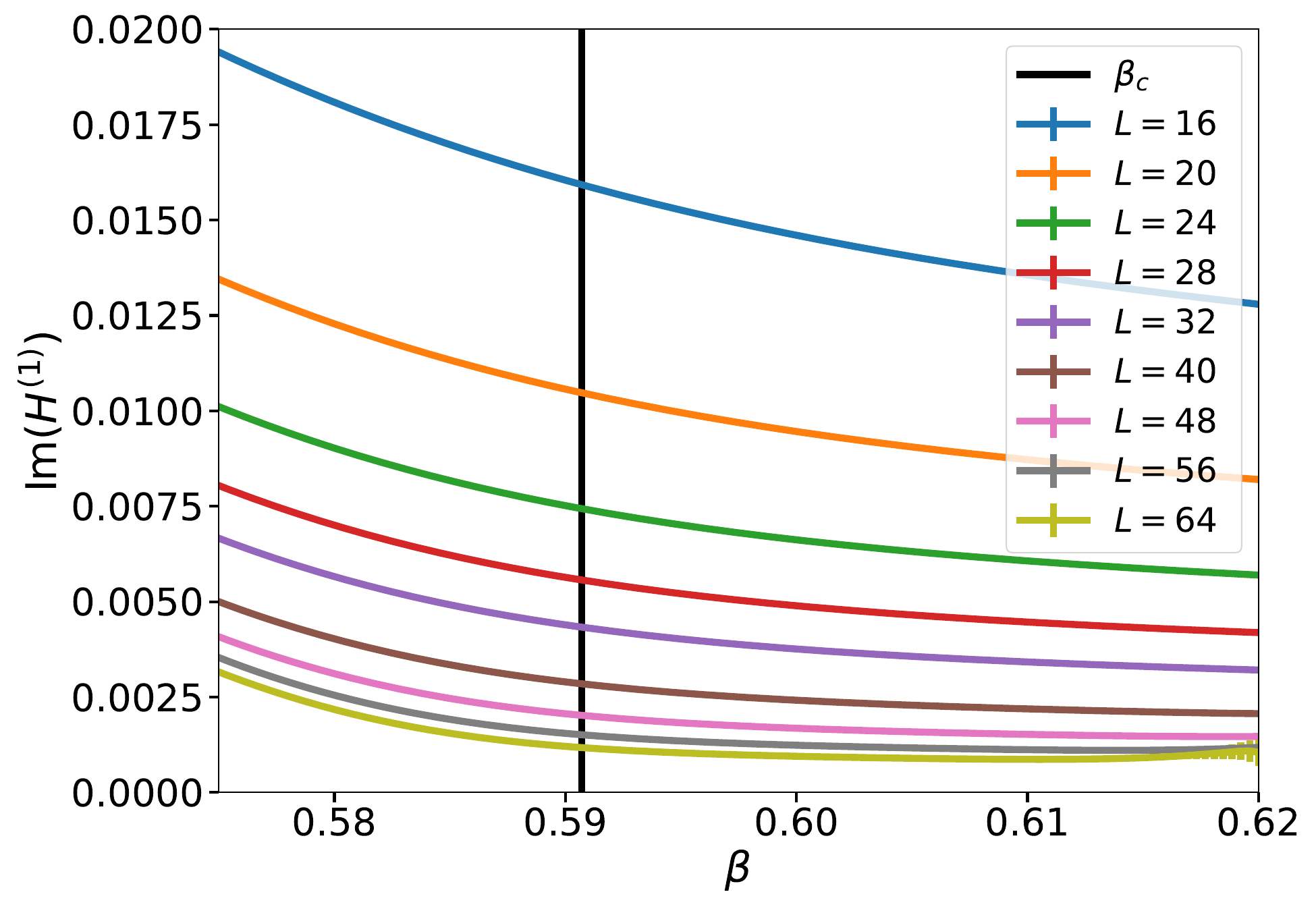}
     \caption{Behaviour of the Lee–Yang zeros as a function of the inverse temperature $\beta$, obtained from the cumulant method method for $n=6$. System sizes: $L=16 - 64$.}
    \label{fig:LY-bplane}
\end{figure}

Finally, it is also interesting to look at the behaviour of the Lee-Yang zeros with respect to the temperature. Figure~\ref{fig:LY-bplane} shows that for $\beta < \beta_{\rm c}$ they are not close to the real axis suggesting that there is a Lee-Yang gap even in the thermodynamic limit, while for $\beta > \beta_{\rm c}$ it seems that the zeros are getting closer and closer to the real axis.

\subsection{Cumulant method}
\label{sec:CM}

In~\cite{deger_determination_2019, deger_lee-yang_2020-1, deger_lee-yang_2020, deger_lee-yang_2018}, Deger and co-authors demonstrated that the first zero of the partition function can be numerically extracted from high-order cumulants of observables--either magnetization or energy-denoted $\langle\!\langle M^n\rangle\!\rangle$ or $\langle\!\langle E^n\rangle\!\rangle$. They showed that, in the limit of large cumulant order $n$, these expressions converge extremely rapidly.

Fluctuations of a quantity, e.g.,~$E$, with respect to a control parameter, e.g.,~$T$ or $\beta$, are defined as derivatives of $\ln Z$: 
\begin{equation}
        \langle\!\langle E^n (\beta)\rangle\!\rangle = (-1)^n \frac{\partial_{\beta}^n \ln {Z}(\beta)}{N}. 
\end{equation}
This relation leads to closed-form expressions for the zeros. For the Fisher zeros, one obtains:
\begin{eqnarray}
     \operatorname{Re} [\beta^{(1)}-\beta] & \approx & 
     \frac{n(n+1) \langle\!\langle E^n(\beta) \rangle\!\rangle \langle\!\langle E^{n+1}(\beta) \rangle\!\rangle - n(n-1) \langle\!\langle E^{n-1}(\beta) \rangle\!\rangle \langle\!\langle E^{n+2}(\beta) \rangle\!\rangle}   
     {2 \left[(n+1) \langle\!\langle E^{n+1}(\beta) \rangle\!\rangle^2 - n \langle\!\langle E^n(\beta) \rangle\!\rangle  \langle\!\langle E^{n+2}(\beta) \rangle\!\rangle\right]}, \label{generalcumulantree}  \\
    |\beta^{(1)}-\beta|^2 & \approx &   
    \frac{n^2(n+1) \langle\!\langle E^n(\beta) \rangle\!\rangle^2  - n(n^2-1) \langle\!\langle E^{n-1}(\beta) \rangle\!\rangle \langle\!\langle E^{n+1}(\beta) \rangle\!\rangle}   
    { (n+1) \langle\!\langle E^{n+1}(\beta) \rangle\!\rangle^2 - n \langle\!\langle E^n(\beta) \rangle\!\rangle  \langle\!\langle E^{n+2}(\beta) \rangle\!\rangle }. \label{generalcumulantimm}
\end{eqnarray}  
Extension to other zeros is obvious.
Formulas~(\ref{generalcumulantree}) and~(\ref{generalcumulantimm}) hold in the asymptotic regime $n \gg 1$ and enable determination of both the real and imaginary parts of the first zero from four consecutive cumulants of $E$, measured at a fixed $\beta$. As emphasized above, this fixed value may lie within a finite window--often broad, especially in the Fisher case--around the critical point $\beta_{\rm c}$.

For the Lee-Yang zeros, odd cumulants $\langle\!\langle M^{2n+1} \rangle\!\rangle$ vanish due to symmetry. Equation~(\ref{generalcumulantimm}) then simplifies to:
\begin{equation}
    \operatorname{Im} [H^{(1)}] \approx \pm \frac{1}{\beta}
    \sqrt{ 2n(2n+1) \left|
    \frac{\langle\!\langle M^{2n} (0)\rangle\!\rangle }
         {\langle\!\langle M^{2(n+1)}(0) \rangle\!\rangle }\right|}.
    \label{eq:cumulantsimp}
\end{equation}
Further discussion can be found in~\cite{moueddenethesis, moueddene_critical_2024}.

\begin{figure}[ht]
    \centering
        \begin{subfigure}[b]{0.9\textwidth}
     \hspace{-1cm}\includegraphics[width=1.05\textwidth]{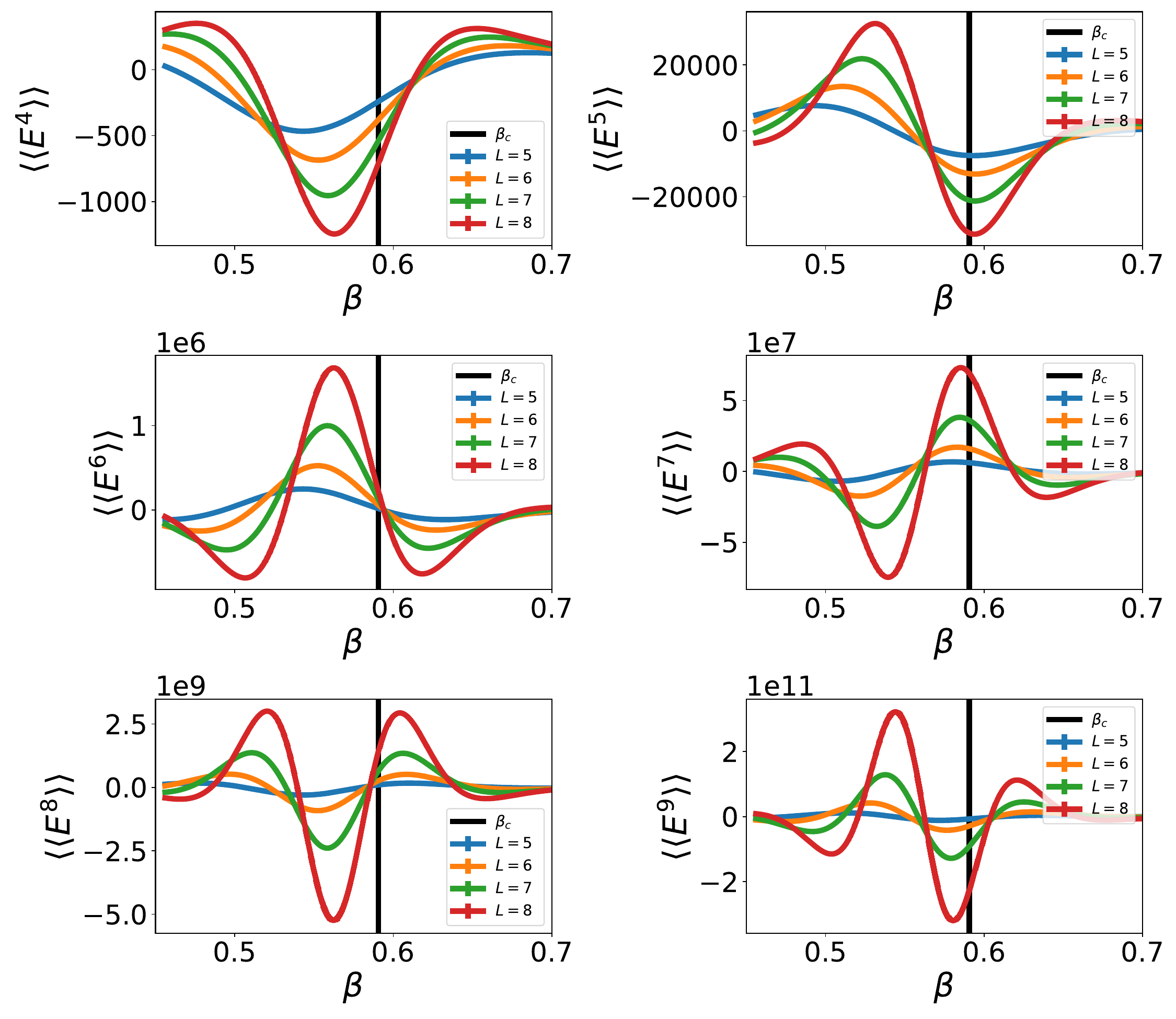}
    \end{subfigure}
  \caption{Even and odd cumulants $\langle\!\langle E^n \rangle\!\rangle$ from $n=4-9$ at $\Delta_{\rm c} = 0$. }
\label{fig:cumulant-allure}
\end{figure}

High-order cumulants are numerically delicate: they tend to vanish rapidly or diverge strongly. It is therefore essential to assess their stability before using them to infer zeros. An illustration of energy cumulants up to $n=9$ is shown in figure~\ref{fig:cumulant-allure}. Two distinct behaviours appear depending on parity: even-order cumulants display a minimum or maximum near $T_{\rm c}$, whereas odd-order cumulants vanish at $T_{\rm c}$. In this chapter, cumulants are obtained from the central moments of the energy, while in~\cite{moueddene_critical_2024-1} they were computed directly from magnetization averages. Small discrepancies can therefore arise; however, the central-moment method consistently yields more stable results.

\begin{figure}[ht]
    \centering
     \includegraphics[width=0.9\textwidth]{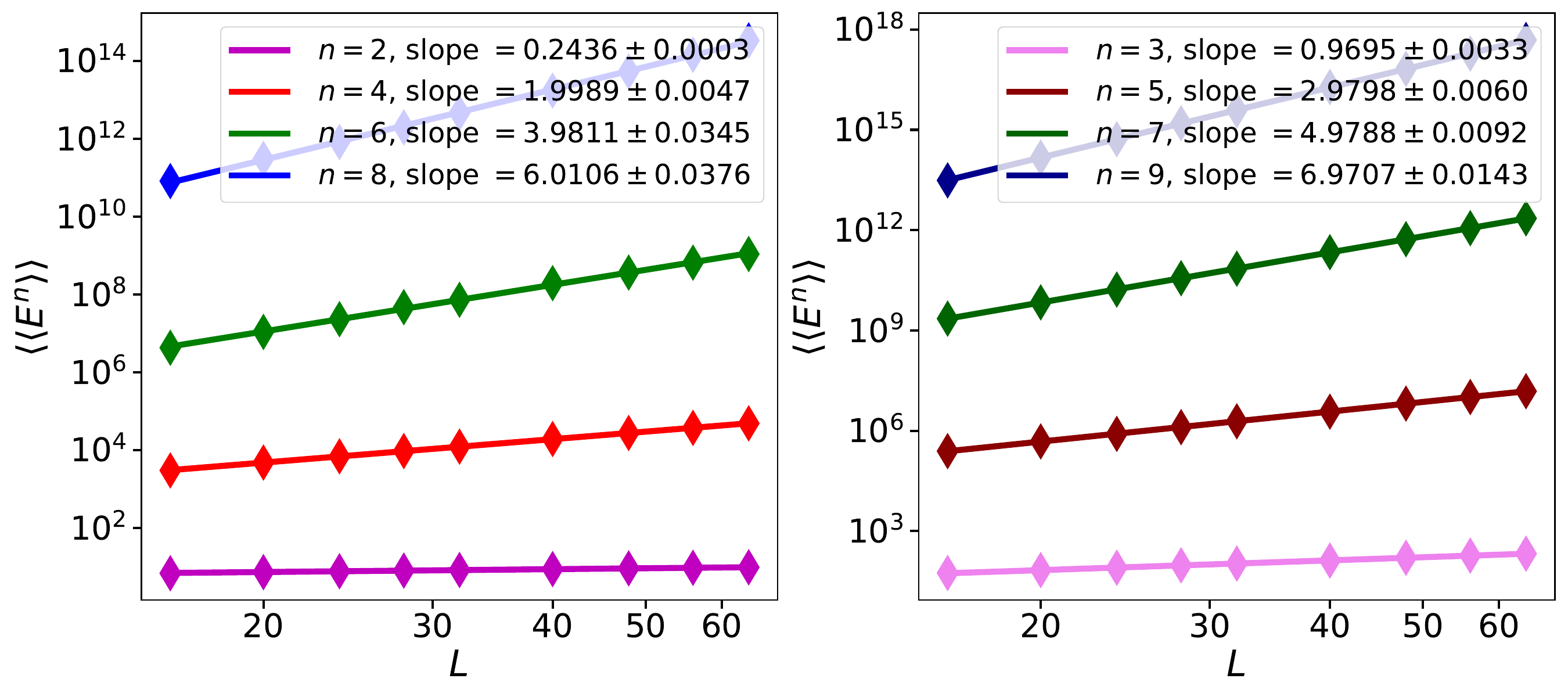}
\caption{Finite-size scaling behaviour of the energy cumulants $\langle\!\langle E^n \rangle\!\rangle$ at the critical point, shown separately for even cumulants (left panel) and odd cumulants (right panel). The data are compared against the expected scaling law $\langle\!\langle E^n \rangle\!\rangle \sim L^{-d + n y^{\rm IM}_t}$. Numerical results for $n = 1 - 9$ are displayed, obtained using the hybrid Metropolis-Wolff algorithm. Both panels use double-logarithmic axes.}
    \label{fig:2DFSS-cumulantsE-Tc}
\end{figure}

Since cumulants can be generated up to large $n$, their internal consistency can be checked via finite-size scaling at criticality. The expected scaling reads:
\begin{equation}
\langle\!\langle E^n\rangle\!\rangle \sim L^{-d+ny_t^{\rm IM}}.
\end{equation}
In two dimensions, this reduces to $\langle\!\langle E^n \rangle\!\rangle \sim L^{\,n - 2}$. On a log-log scale, this implies a straight line with slope $n - 2$, as confirmed in figure~\ref{fig:2DFSS-cumulantsE-Tc}. 
Separate fits for even and odd cumulants show acceptable convergence even for $n \geq 1$, with particularly precise agreement for even cumulants when $n \geq 2$.

The close match between measured slopes and the theoretical expectation validates the cumulant approach: high-order cumulants from relatively small lattices already reproduce the correct scaling behaviour. Their rapid convergence provides accurate estimates of the first Fisher zero, while exponential convergence with order has been confirmed in other models (Ising, Potts), attesting to the robustness of the cumulant method. This has been exploited in tables 5 and 6 in \cite{moueddene_critical_2024} for both Fisher and Lee-Yang zeros.

\section{Tricritical point singularities}
\label{sec:tp-sing}

\subsection{Off-critical scaling}
\label{sec:off-crit-scal}

\begin{figure}[H]
    \centering
     \includegraphics[width=0.45\textwidth]{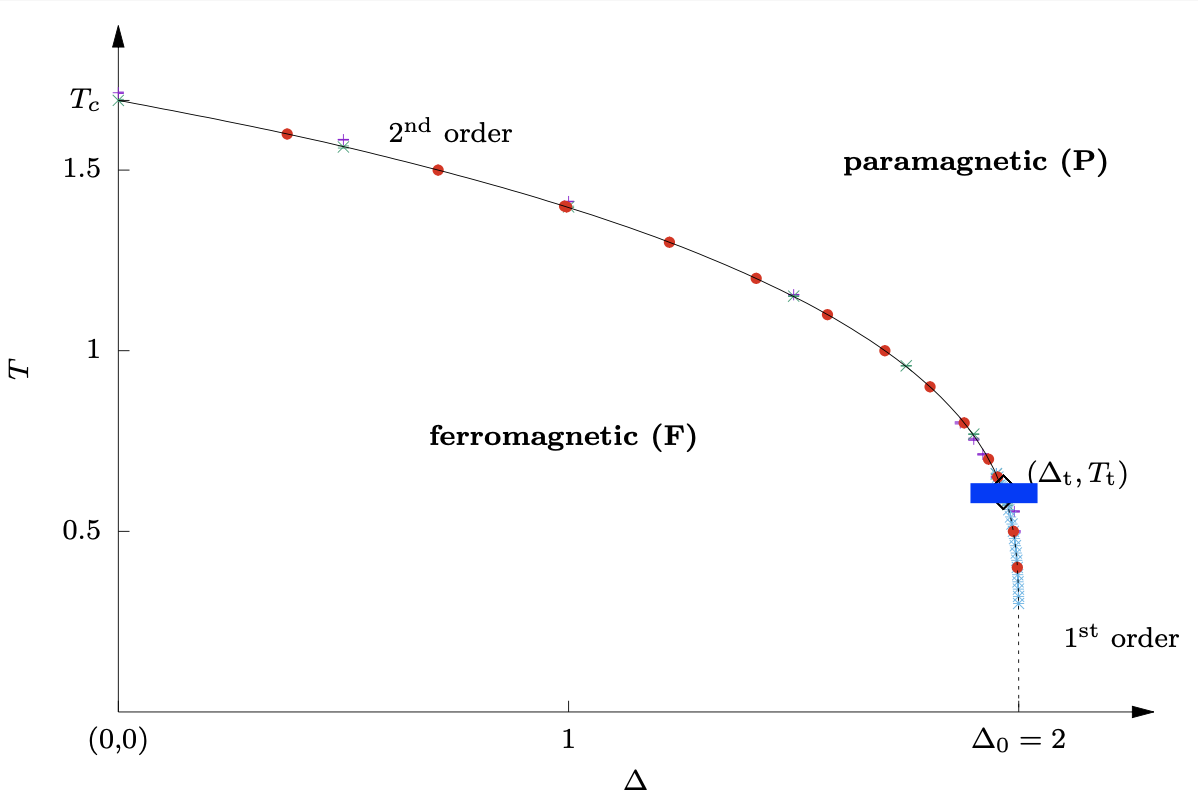}
\includegraphics[width=0.45\textwidth]{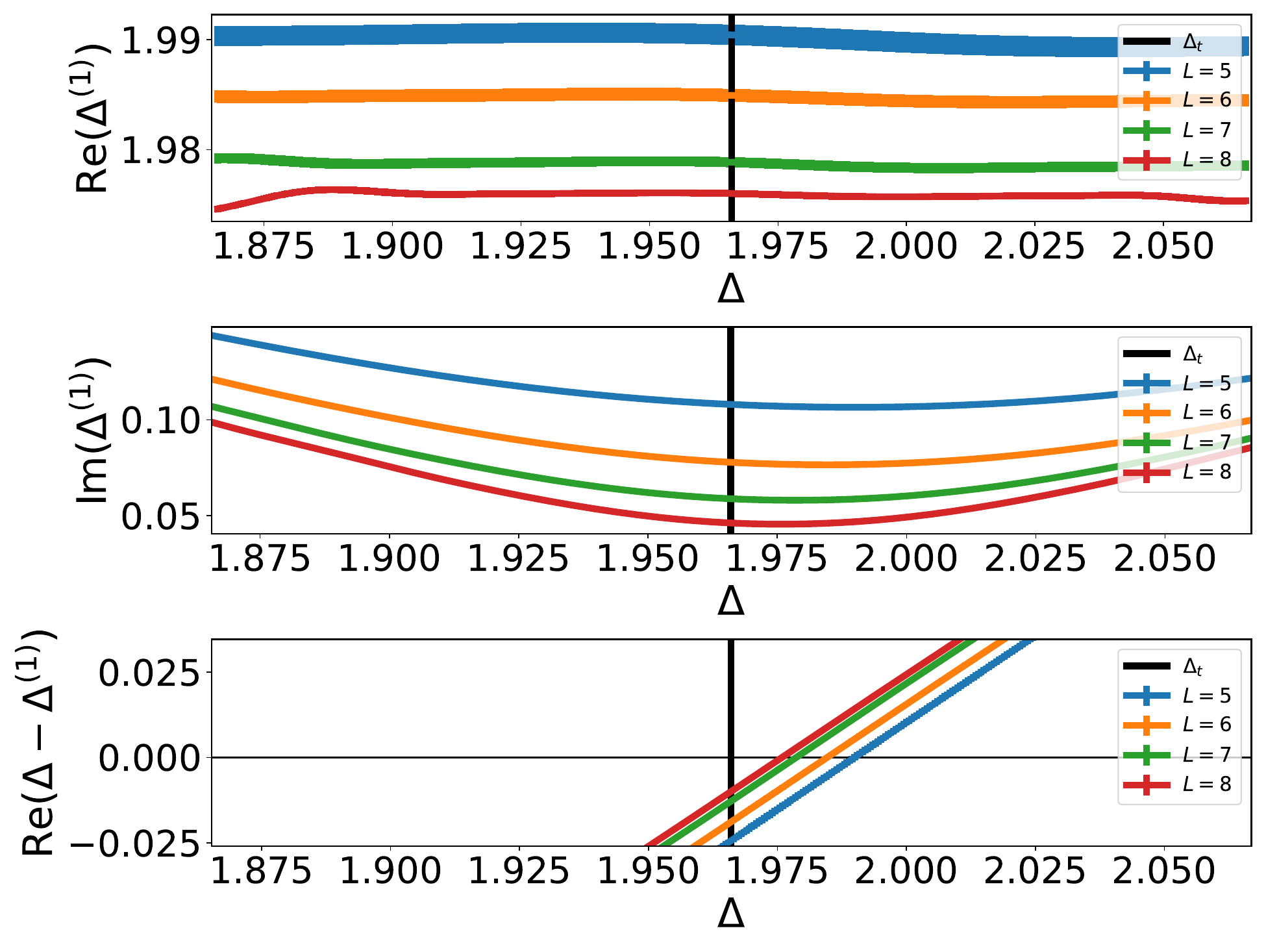}
\caption{Variation of the real and imaginary parts of $\Delta^{(1)}$, extracted via the cumulant method, as a function of $\Delta$. System sizes: $L=16 - 64$. Solid lines represent $\operatorname{Re} (\Delta - \Delta_r^{(1)})$, while dashed lines correspond to $\operatorname{Im}(\Delta^{(1)})$. Vertical dotted lines are included as a visual guide to indicate the imaginary part associated with the vanishing of $\operatorname{Re} (\Delta - \Delta^{(1)})$.}
\label{fig:cumulantdelta}
\end{figure}

In this section, we investigate the stability range of the first zero near the tricritical point. In this regime, it is more appropriate to analyse crystal-field zeros rather than Fisher zeros, since the transition line at $(T_{\rm t}, \Delta_{\rm t})$ is oriented nearly perpendicular to the crystal-field axis in the phase diagram (see figure~\ref{fig:cumulantdelta}). Accordingly, the crystal field $\Delta$ serves as a natural parameter for probing the transition. In the right panel of the figure, we first locate the region where the real part $\operatorname{Re} (\Delta - \Delta^{(1)})$ vanishes, and then examine both $\operatorname{Re} (\Delta^{(1)})$ and $\operatorname{Im} (\Delta^{(1)})$ as functions of $\Delta$. Although the stability observed here is less pronounced than in the case of the Fisher zero at $\Delta_{\rm c} = 0$, the first crystal-field zero nevertheless remains remarkably stable within the interval $0.95,\Delta_{\rm t} \leq \Delta \leq 1.05,\Delta_{\rm t}$, which represents a fairly broad window.

\subsection{Cumulant scaling}
\label{sec:cum_scal}

\begin{figure}[ht]
  \centering
    \includegraphics[width=0.9\linewidth]{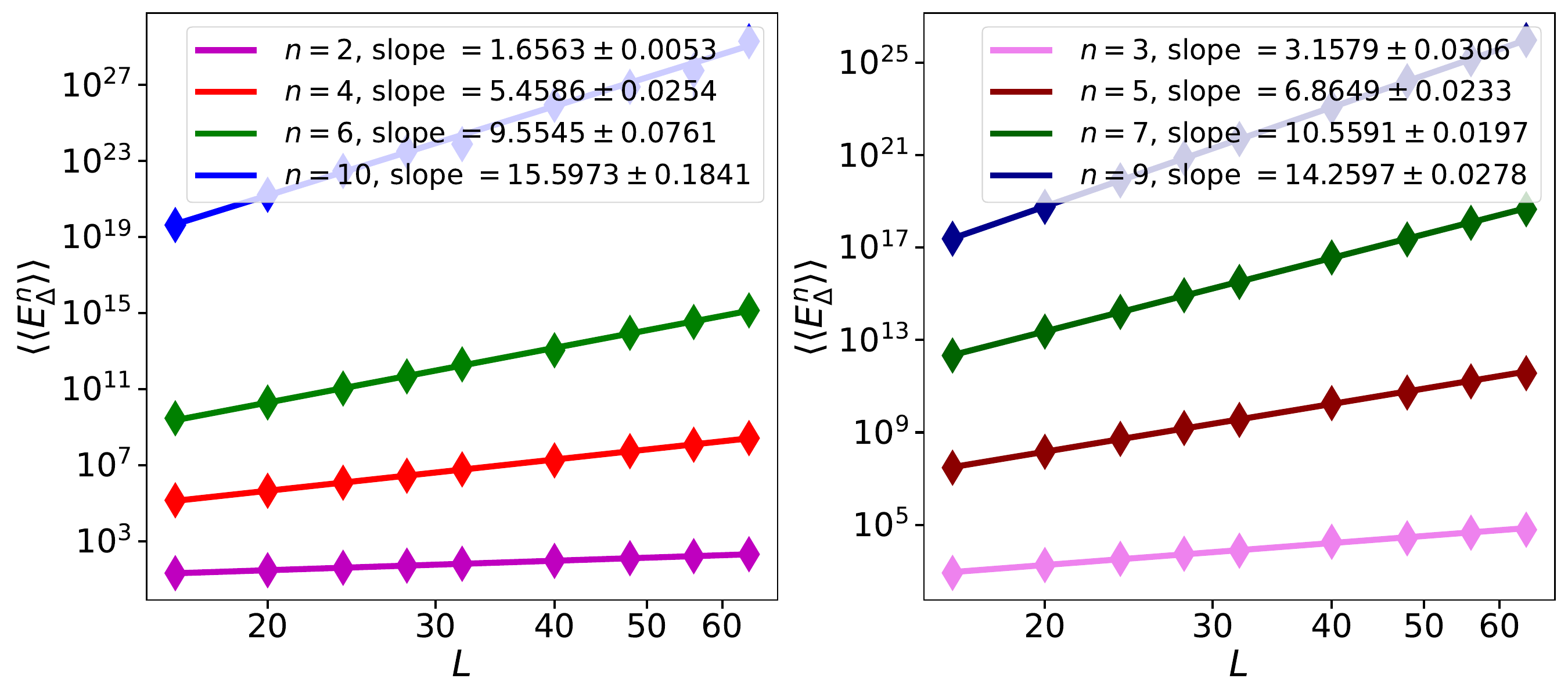} 
    \caption{Finite-size scaling behaviour of the crystal-field energy cumulants $\langle\!\langle E_\Delta^n \rangle\!\rangle$ at the tricritical point. The left panel shows even-order cumulants, and the right panel shows odd-order cumulants, following the scaling law $\langle\!\langle E_\Delta^n \rangle\!\rangle \sim L^{-d + n y_t^{\rm TP}}$. Numerical results, obtained via the hybrid Metropolis-Wolff protocol, are shown for $n = 2 - 10$. Note the double-logarithmic scale.}
    \label{fig:cumulantEd}
\end{figure}

As for the case $\Delta_{\rm c} = 0$, we now focus on the tricritical point by extracting the thermal exponent from the scaling of the cumulants of the $\Delta$-contribution to the energy, defined as
\begin{equation}
    E_\Delta = \Delta \sum_i \sigma_i^2.
\end{equation}
The results of the cumulant scaling are presented in figure~\ref{fig:cumulantEd}. Compared to the critical case, the exponents--both odd and even--deviate more noticeably from the expected tricritical values, suggesting that either corrections to scaling must be included or larger system sizes are required. The goodness of fit, measured by the reduced chi-square, is approximately $\chi^2_s \approx 2$ in each case.

\subsection{First crystal-field zero scaling}
\label{sec:first_CF_zero}

\begin{figure}[H]
    \centering
     \includegraphics[width=0.6\textwidth]{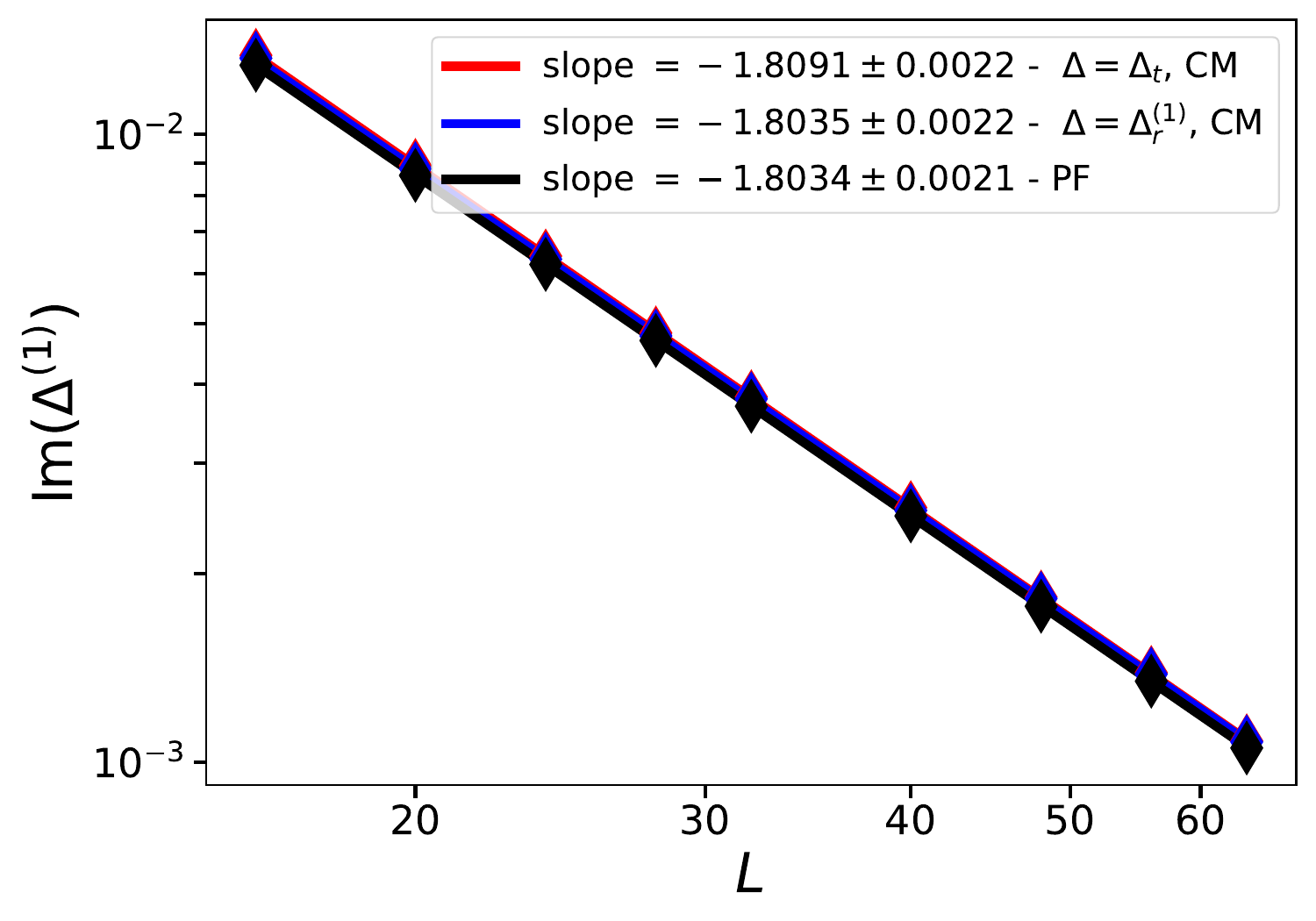}   \caption{Finite-size scaling of the imaginary part of the first crystal-field zero, $\operatorname{Im} (\Delta^{(1)})$. Results are obtained via the cumulant method (CM), evaluated at $\Delta = \Delta_{\rm t}$ and at $\Delta = \Delta_r^{(1)}$, and compared with the partition function (PF) method at $T = 0.608$.}
\label{fig:ImdeltaCM}
\end{figure}

We now turn to the finite-size scaling analysis of the real and imaginary parts of the first crystal-field zero, determined either directly from the partition function or via the cumulant expansion. Figure~\ref{fig:ImdeltaCM} presents the scaling of the imaginary part $\operatorname{Im}(\Delta^{(1)})$, evaluated at both $\Delta=\Delta_{\rm t}$ and $\Delta=\Delta_r^{(1)}$, and compared against the partition function method for system sizes $L=16$–$64$. The agreement is excellent: at $\Delta = \Delta_r^{(1)}$ we obtain $y_t = 1.8035(22)$ with $\chi^2_s = 2.2$, in almost perfect correspondence with the partition function result $y_t = 1.8034(21)$. At $\Delta=\Delta_{\rm t}$ the fit yields $y_t = 1.8091(22)$ with $\chi^2_s = 1.51$, which is likewise consistent within statistical uncertainties. 

\begin{figure}[H]
    \centering
     \includegraphics[width=0.6\textwidth]{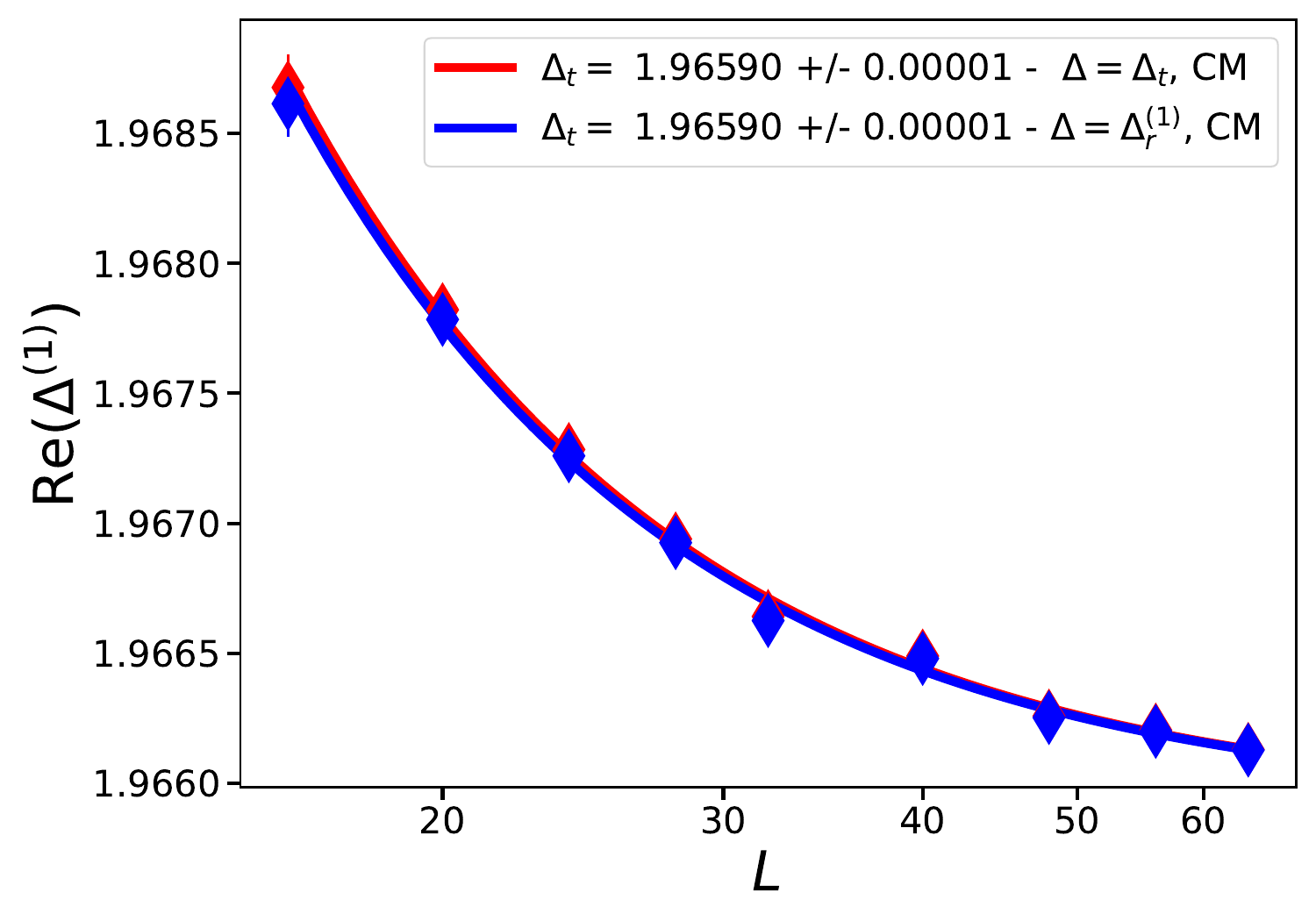} 
     \caption{Finite-size scaling behaviour of the real part of the first crystal-field zero, $\operatorname{Re}(\Delta^{(1)})$. Results are obtained via the cumulant method (CM), evaluated at $\Delta = \Delta_{\rm t}$ and at $\Delta = \Delta_r^{(1)}$, and used to extract $\Delta_{\rm t}$.}
\label{fig:RedeltaCM}
\end{figure}

For the real part, the results are presented in figure~\ref{fig:RedeltaCM}. Unlike the imaginary component, fitting the scaling of $\operatorname{Re}(\Delta^{(1)})$ with $\Delta_{\rm t}$ fixed yields a poor estimate of the exponent $y_t$. However, when the tricritical exponent is fixed to $y_t^{\rm TP} = 1.80$, the shift behaviour of the real part provides an excellent determination of $\Delta_{\rm t}$. Specifically, we obtain $\Delta_{\rm t} = 1.96590(1)$ with $\chi^2_s = 1.94$ at $\Delta=\Delta_{\rm t}$, and $\Delta_{\rm t} = 1.96590(1)$ with $\chi^2_s = 1.95$ at $\Delta=\Delta_r^{(1)}$, in remarkable agreement between the two approaches.

\subsection{First Lee-Yang zero scaling}
\label{sec:first_LY_zero}

\begin{figure}[H]
  \centering
    \includegraphics[width=0.57\linewidth]{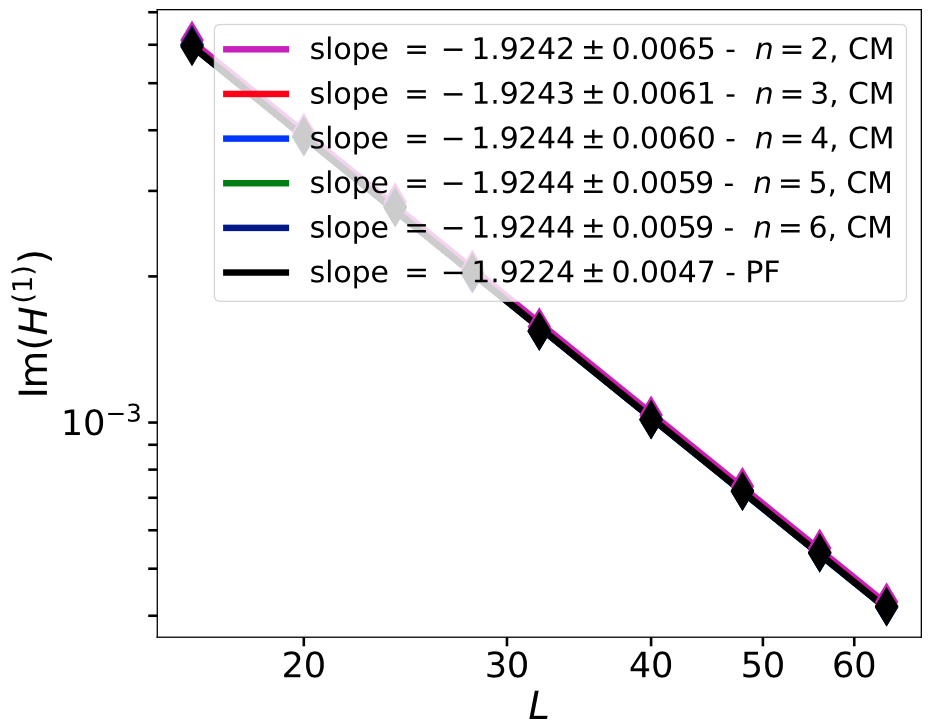}
    \caption{Finite-size scaling behaviour of the Lee–Yang zeros at the  tricritical  point. Comparative results from the cumulant method (CM) and partition function (PF) method are shown. The expected value is $y_h^{\rm TP} = 1.925$.}
\label{fig:FSSHn}
\end{figure}

The finite-size scaling analysis of the first Lee–Yang zero, obtained via the cumulant method up to order $n=3$, has been extensively studied in~\cite{moueddene_critical_2024}. In the present work, we extend this analysis to $n=6$ and demonstrate that the convergence continues to improve with higher-order cumulants; see figure~\ref{fig:FSSHn}.

\section{Density of the first zeros}
\label{sec:dens_first_zeros}

The density of partition function zeros provides valuable insight into both the nature and strength of phase transitions. This line of investigation was systematically pursued by Janke and Kenna~\cite{janke_analysis_2001}, who demonstrated its effectiveness across a wide range of systems, even when only relatively small lattice sizes were available. 
Subsequently, Alves \emph{et al}~\cite{Alves_2002} compared this classification scheme with an alternative approach introduced by Borrmann \emph{et al}~\cite{borrmann_classification_2000}, which relies on the linear scaling of the limiting density of zeros. Their analysis showed that finite-size scaling of modest system sizes is already sufficient to distinguish the order of the transition in the $4$- and $5$-state Potts models. More recently, the density of Lee-Yang zeros in the three-dimensional Blume–Capel model was investigated in~\cite{moueddene_critical_2024-1}.

In this framework, the cumulative distribution of zeros can be written as
\begin{equation}
G_L\bigl(r^{(j)}_i\bigr) = a_1  r^{(j)}_i L^{a_2} + a_3 ,
\end{equation}
which encapsulates key information about the phase transition. At the transition, the constant term $a_3$ must vanish, while the exponent $a_2$ determines the nature of the transition. For a first-order transition one expects $a_2 \approx 1$, with the prefactor $a_1$ directly related to the latent heat $\Delta e$ (for Fisher zeros) or to the spontaneous magnetisation (for Lee–Yang zeros). By contrast, for a second-order transition one finds $a_2 = 2 - \alpha = d/y_t$ for Fisher zeros, and $a_2 = d/y_h$ for Lee–Yang zeros. Here, $r^{(j)}_i$ denotes the imaginary part of the $j$-th zero.

At the critical point $\Delta_{\rm c} = 0$, the finite-size scaling of the Fisher zero density is analysed through
\begin{equation}
G_L(\beta_i^{(1)})=a_1  \beta_i^{(1)}L^{\frac{d}{y_t}}+a_3 ,
\end{equation}
whereas at the tricritical point the analysis focuses on the scaling of the crystal-field zeros:
\begin{equation}
G_L(\Delta_i^{(1)})=a_1  \Delta_i^{(1)}L^{\frac{d}{y_t}}+a_3 .
\end{equation}

\begin{figure}[H]
    \centering
    \includegraphics[width=1.05\textwidth]{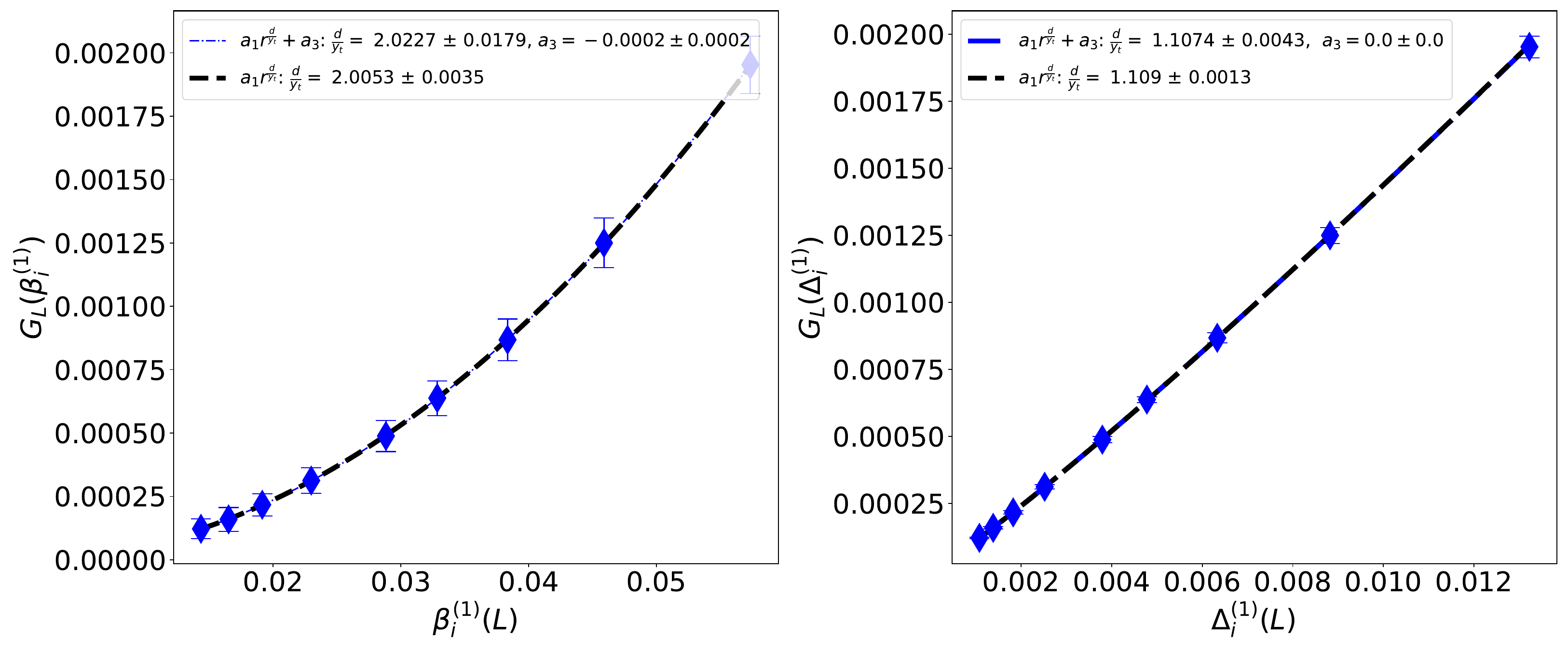}
    \caption{\textbf{Left panel:} Finite-size scaling analysis of the density of the imaginary part of the first Fisher zero, $\beta_i^{(1)}$, at $\Delta=0$. The expected exponent is $d/y_t^{\rm IM} = 2$. \textbf{Right panel:} Finite-size scaling analysis of the density of the first crystal-field zero at $T_{\rm t} = 0.608$. The expected exponent is $d/y_t^{\rm TP} \approx 1.111$.}
\label{fig:density2dyt}
\end{figure}

This latter case was first investigated in~\cite{moueddenethesis}. Figure~\ref{fig:density2dyt} shows the Fisher zero density at the critical point (left panel) and the crystal-field zero density at the tricritical point (right panel), for system sizes $L = 16 - 64$. In both cases, we perform two separate fits. In the first, $a_3$ is kept as a free parameter to test whether it vanishes--a necessary condition for the presence of a phase transition. 
For both the critical and tricritical analyses, the fitted values of $a_3$ are indeed consistent with zero. Having established this, we then set $a_3 = 0$ in a second round of fits, which allows us to extract the critical exponents with improved accuracy. 

At the critical point, the fit yields $d/y_t = 2.0053(35)$, in excellent agreement with the exact value $d/y_t^{\text{IM}} = 2$. At the tricritical point, where the theoretical prediction is $d/y_t^{\text{tri}} \approx 1.111$, we obtain the numerical estimate $d/y_t = 1.1090(13)$, again in strikingly good agreement. 

At the tricritical point, we extend previous analyses by investigating crossover effects in its vicinity ($\Delta_{\rm t} = 1.966$, $T_{\rm t} = 0.608$) through the behaviour of the Lee–Yang zeros. As demonstrated in~\cite{moueddene_critical_2024, moueddene_critical_2024-1}, Lee–Yang zeros are highly sensitive to variations in the external parameters, making them a powerful probe of the three distinct regimes expected in the Blume–Capel phase diagram near tricriticality. To explore this sensitivity, we analyse two nearby temperatures--one slightly below the tricritical point, $T = 0.6075$, and one slightly above, $T = 0.6083$--using the same parameter values employed in the analysis of figure 8 in~\cite{moueddene_critical_2024}. In all cases, the crystal-field parameter is fixed to its tricritical value $\Delta_{\rm t}$.

\begin{figure}[H]
    \centering
     \includegraphics[width=1.05\textwidth]{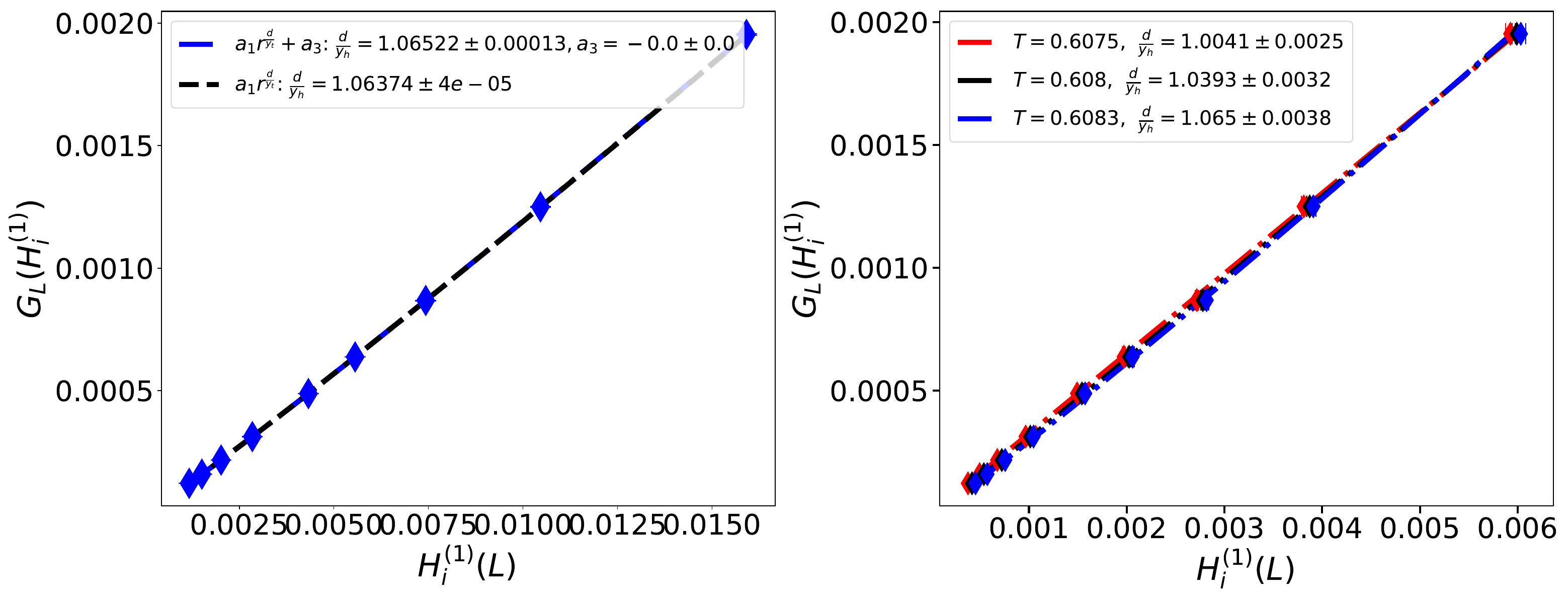}
     \caption{\textbf{Left panel:} Finite-size scaling analysis of the density of the imaginary part of the first Lee–Yang zero, $h_i^{(1)}$, at the critical temperature $T_{\rm c} = 1.6929$. The expected exponent is $d/y_h^{\rm IM} \approx 1.0666$. \textbf{Right panel:} Finite-size scaling  analysis of the density of the first Lee–Yang zero at various temperatures. The expected exponent is $d/y_h^{\rm TP} \approx 1.03896$.}
\label{fig:density2dyh}
\end{figure} 

As in the previous analyses, the fitting procedure starts by checking whether $a_3$ vanishes, following the approach illustrated in figure~\ref{fig:density2dyh} at criticality. Once this condition is confirmed, we fix $a_3 = 0$ to obtain a more precise estimate of the critical exponent. At the critical point, this procedure yields $d/y_h = 1.0637(2)$, in excellent agreement with the Ising universality class prediction, $d/y_h^{\text{IM}} \approx 1.0666$.
At the tricritical point, the estimate $d/y_h = 1.0393(32)$ closely matches the theoretical value $d/y_h^{\text{tri}} \approx 1.03896$.  
Examining nearby temperatures reveals crossover effects: for $T = 0.6075$, the fitted exponent $d/y_h = 1.0041(25)$ clearly indicates a first-order transition, consistent with the expected value $d/y_h = 1$. Conversely, at $T = 0.6083$, the extracted value $d/y_h = 1.0650(38)$ aligns remarkably well with the Ising universality class, highlighting the sensitivity of the Lee–Yang zeros to the external parameters and the transition between different scaling regimes near tricriticality.

\begin{figure}[H]
    \centering
     \includegraphics[width=1.05\textwidth]{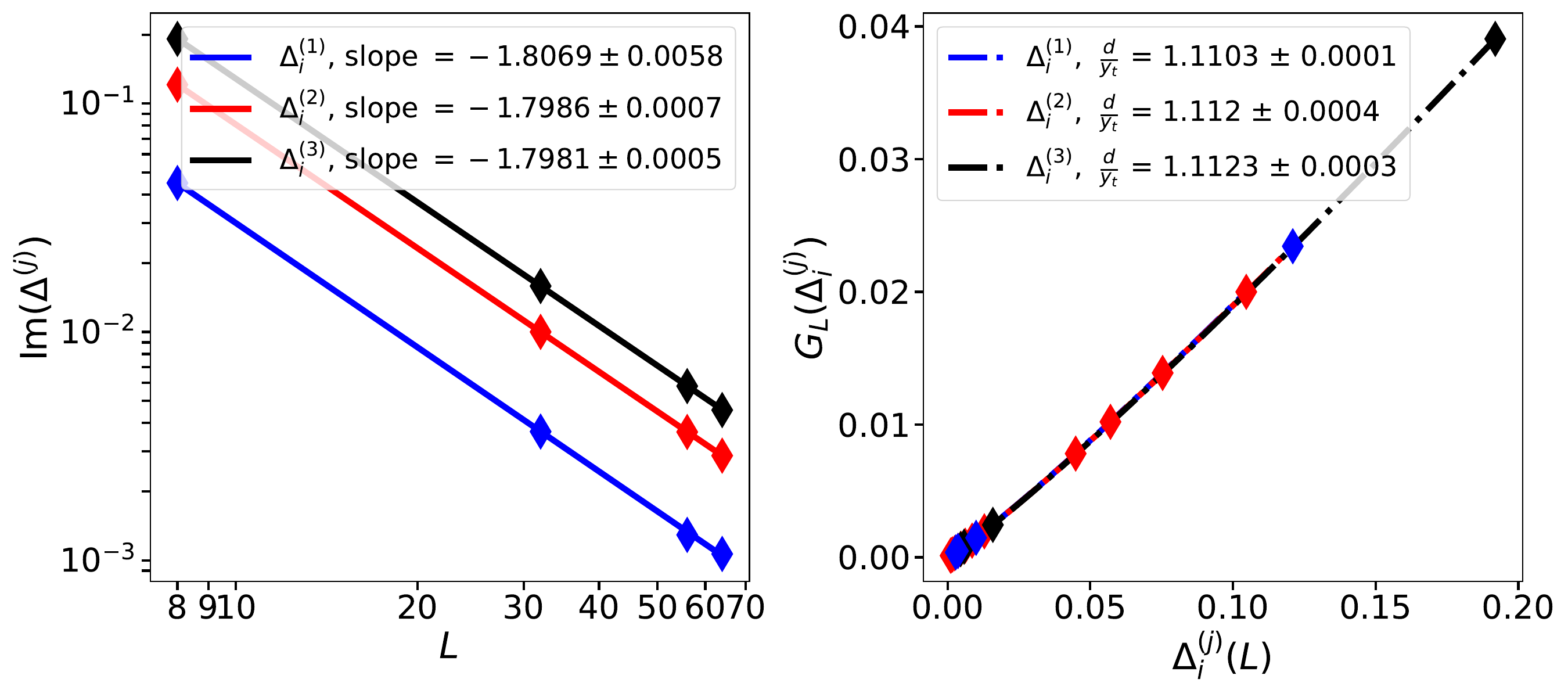}
\caption{\textbf{Left panel:} Finite-size scaling analysis of the first, second, and third crystal-field zeros, $\Delta_i^{(j)}$, at the tricritical temperature $T_{\rm t} = 0.608$. The expected exponent is $d/y_t^{\rm TP} = 1.80$. \textbf{Right panel:} Finite-size scaling analysis of the density of the imaginary parts of the first, second, and third crystal-field zeros, $\Delta_i^{(j)}$. The expected exponent is $d/y_t^{\rm TP} \approx 1.111$.}
\label{fig:density2dytcf}
\end{figure}

Finally, we extend the analysis beyond the leading zero to include higher-order zeros, which also carry valuable information on the critical behaviour of the system~\cite{Ruiz-Lorenzo:2024jwf}. While the cumulant method provides the leading zero, the higher-order zeros were extracted by tracking sign changes of the partition function in the complex crystal-field plane. Our study focuses on the tricritical point for system sizes $L = 8$, $32$, $56$, and $64$, where the first three crystal-field zeros were determined. Their finite-size scaling behaviour is presented in the left panel of figure~\ref{fig:density2dytcf}, alongside the corresponding density of crystal-field zeros in the right panel. Because the zeros are correlated at each lattice size, independent fits were performed for each zero.

The analysis of the second and third crystal-field zeros provides highly consistent estimates of the tricritical exponent, yielding $y_t^{\rm TP} = 1.7986(7)$ from $\Delta_i^{(2)}$ and $y_t^{\rm TP} = 1.7981(5)$ from $\Delta_i^{(3)}$. Likewise, the corresponding density-of-zeros analysis, shown in figure~\ref{fig:density2dytcf} (right panel), produces results in excellent agreement with the literature value $d/y_t^{\rm TP} \approx 1.11$, specifically $d/y_t = 1.1120(4)$ for the second zero and $d/y_t = 1.1123(3)$ for the third zero.

\section{Conclusions}
\label{sec:conclusions}

In this work, we have extended the study of partition function zeros as a tool for analyzing critical and tricritical behaviour in the two-dimensional Blume–Capel model. Building on our previous studies~\cite{moueddene_critical_2024,moueddenethesis}, we examined Lee–Yang zeros in the complex magnetic-field plane, Fisher zeros in the complex-temperature plane, and crystal-field zeros, systematically exploring their finite-size scaling properties along both the second-order critical line and near the tricritical point. Our results show that accurate estimates of critical exponents can be obtained even from simulations performed away from the nominal transition points, thereby confirming and generalizing the findings of Deger \emph{et al}~\cite{deger_determination_2019, deger_lee-yang_2020-1}. Notably, reliable results are achievable with surprisingly modest lattice sizes, significantly reducing the computational effort required.

Overall, our study highlights the versatility and efficiency of the partition function zero approach, especially when combined with the cumulant method. By showing that reliable critical information can be extracted from small systems and off-critical simulations, this work not only confirms theoretical predictions for the Blume–Capel model but also provides a framework for computationally efficient investigations of other complex systems.

\section*{Acknowledgments}
The work of N G F was supported by the  Engineering and Physical Sciences Research Council (grant EP/X026116/1 is acknowledged).

\printbibliography

\end{document}